\documentclass[pre,twocolumn,aps]{revtex4}

\usepackage{amsmath,amssymb}
\usepackage{latexsym}
\usepackage{graphicx}
\usepackage{color}

\def\rootfig{./}

\begin{document}
\title{Exploring Rigidly Rotating Vortex Configurations and their Bifurcations
in Atomic Bose-Einstein Condensates}

\author{
A.V.\ Zampetaki$^1$,
R.\ Carretero-Gonz\'{a}lez$^2$,
P.G.\ Kevrekidis$^3$, F.K. Diakonos$^4$
and D.J. Frantzeskakis$^4$}

\affiliation{%
$^1$
Zentrum f\"{u}r Optische Quantentechnologien,
Universit\"{a}t Hamburg, Luruper Chaussee 149, 22761 Hamburg, Germany
\\
$^2$Nonlinear Dynamical Systems Group,
Computational Science Research Center, and
Department of Mathematics and Statistics,
San Diego State University, San Diego,
CA 92182-7720, USA
\\
$^3$ Department of Mathematics and Statistics, University of
Massachusetts, Amherst MA 01003-4515, USA
\\
$^4$ Department of Physics, University of Athens, Panepistimiopolis,
Zografos, Athens 157 84, Greece
}

\begin{abstract}
In the present work, we consider the problem of a system of few vortices
$N \leq 5$ as it emerges from its experimental realization in the
field of atomic Bose-Einstein condensates. Starting from the corresponding
equations of motion
for an axially symmetric trapped condensate,
we use a two-pronged approach in order to reveal the
configuration space of the system's preferred dynamical states. On the one
hand, we use a Monte-Carlo method parametrizing the vortex ``particles''
by means of hyperspherical coordinates and identifying the minimal energy
ground states thereof for $N=2, \dots, 5$ and different vortex
particle angular momenta. We then complement this picture
with a dynamical systems analysis of the possible rigidly rotating states.
The latter reveals supercritical and subcritical pitchfork, as well
as saddle-center bifurcations that arise exposing the full wealth of the
problem even at such low dimensional cases. By corroborating the results
of the two methods, it becomes fairly transparent which branch the Monte-Carlo
approach selects for different values of the angular momentum which is used
as a bifurcation parameter.
\end{abstract}

\maketitle

\section{Introduction}

Over the past fifteen years, there has been an intense interest on the
dynamics of nonlinear waves and coherent structures that arise in
the atomic physics realm of Bose-Einstein
condensates~\cite{pethick,stringari,emergent}. A large component
to the appeal of such states has been the apparent simplicity
and controllability of this setting, which at the mean field level
can be well approximated by the so-called Gross-Pitaevskii equation
where such states as solitary waves and vortices have been
widely explored~\cite{emergent}. Among the relevant
coherent structures, vortices have, arguably, held a prominent
position, perhaps in part due to the tantalizing analogies to
earlier studies of their existence in fluids; relevant research
activity focusing on vortices has now been summarized in
multiple works~\cite{donnely,fetter1,fetter2,usmplb,tsubota,brianrl}.

Some of the early interest in vortex structures has been
centered around their experimental realization in various
distinct ways. Additionally, vortices of higher
topological charge were produced and their decay
was explored~\cite{middel16}. Finally, large scale lattices
featuring triangular symmetry were demonstrated as the emerging
ground state of the system under fast rotation~\cite{middel13}.
After what could be considered as a partial experimental research
hiatus in the middle of the last decade, a series of recently
devised techniques shifted the interests within the paradigm
of vortices in BECs and gave rise to new possibilities
accessible both in their creation and in their monitoring.
In particular, the possibility to create the vortices
by quenching through the condensation quantum phase
transition~\cite{BPA08} was coupled to minimally
destructive imaging techniques~\cite{freilich10} and enabled
the visualization of single vortex precessions but
also multi-vortex interactions. The latter included
both the case of the counter-circulating vortex
dipole~\cite{BPA10,freilich10,middel_pra11}, but
also more recently that of co-rotating sets of
$N=2, \dots, 5$ vortices~\cite{corot_prl}. The case of $N=3$ was
also explored through different experimental techniques,
involving the excitation of a quadrupolar mode in Ref.~\cite{bagn}.
While such few vortex clusters were created early
on in the experimental history of atomic BECs~\cite{middel8}
and were intensely studied theoretically~\cite{castindum,middel25,middel26,middel27,middel28,middel29,middel30,middel31,middel32,middel2010}, these
recent works have shed new light into relevant static and dynamic
possibilities that has, in turn, motivated further theoretical
analysis~\cite{middel24,middel_physd,middel_pla,middel_anis}.

Our principal aim in the present work is to revisit and
expand upon the recent experimental, computational
and theoretical discussion of Ref.~\cite{corot_prl}. Given that
the latter work offers a well established framework of
ordinary differential equations for tracking the vortex motion, here
we wish to advance to the extent possible our state of
understanding of such low dimensional reductions of the system
by employing a variety of computational and theoretical tools.
In particular, on the computational side, we interweave two
different approaches. On the one hand, we use
a Monte-Carlo (MC) based technique involving a
twist of a reparametrization method for the vortex ``particles''
based on hyperspherical coordinates. This approach will prove
extremely efficient in unveiling the ground state of the system.
On the other hand, we use the computational software package
AUTO~\cite{auto} in order to provide a the bifurcation picture for
the cases of $N=2$, $N=3$ and $N=4$. The combination of the
two methods sheds further light on the parameter values
(in this case, the angular momentum as discussed below) for which
 the MC jumps from one
type of solution to a next. We corroborate these results by
systematic analytical results on the co-rotating vortex states,
whereby we explore not only the stability of the most
standard polygonal state~\cite{castindum}, but also explore
by-products of the bifurcations thereof. In the case of
$N=2$, these are asymmetrically located anti-diametric
vortices, for $N=3$, they form isosceles as opposed to
equilateral triangles, for $N=4$, rhombi emerge instead of
squares, etc.

Our presentation is structured as follows.
In section II, we present the basic equations associated
with the vortex dynamics and their mathematical framework (conserved
quantities etc.). In section III, we analyze the Monte-Carlo
method used to address the ground state solutions of this system of equations.
In section IV, we present our numerical and analytical results, separating
the cases of $N=2$, $N=3$, $N=4$ and even briefly touching upon $N=5$.
Finally, in section V, we summarize our findings and present our conclusions
as well as a number of directions of interest for future studies.

\section{The Equations of Motion and Conservation Laws}
For the recent experimental results of Ref.~\cite{corot_prl}, it was argued
by a combination of numerical and theoretical results
(see for related recent analyses also the works of
Refs.~\cite{chang,pelin}) that the dynamics for
$N$ singled-charged BEC vortices trapped in an
axially symmetric
magnetic trap
can be described by the following system of differential equations:
\begin{eqnarray}
\label{ODEs}
\dot{r}_{i} &=&
\displaystyle
-c\, \sum_{i \neq j} S_j
\frac{r_j}{\rho_{ij}^2} \sin(\theta_i-\theta_j),
\\[0.5ex]
\notag
\dot{\theta}_{i} &=&
\displaystyle
\frac{S_{i}}{1-r_i^2}
+  c\, \sum_{i \neq j} S_j \left[
\frac{1}{\rho_{ij}^2} -
\frac{r_j}{r_i\,\rho_{ij}^2} \cos(\theta_i-\theta_j) \right],
\end{eqnarray}
where $S_i$ is the charge of vortex $i$ and its position, rescaled by the
Thomas-Fermi (TF) cloud
radius $R_{\rm TF}$, is given in polar coordinates by $(r_i,\theta_i)$,
$\rho_{ij}$ is the distance between vortices $i$ and $j$, and
 $c=\frac{1}{2}({\omega_{\rm vor}}/{\omega_{\rm pr}^{0}})$
is an adimensional parameter accounting for the ratio of the rotation
frequency of two same charge vortices ($\omega_{\rm vort}/d^2$ when
the vortices are separated by a distance $d$ again measured in
units of $R_{\rm TF}$) and the rotational precession
induced by the magnetic trap.
The precession of a single vortex about the trap center can be approximated by
$\omega_{\rm pr}={\omega_{\rm pr}^{0}}/({1-{r^{2}}})$,
where
the frequency at the trap center is
$\omega _{\rm pr}^{0}=\ln \left( A\frac{\mu }{\Omega }\right)/R_{\rm TF}^{2}$,
$\mu$ is the chemical potential, and $A$ is
a numerical constant~\cite{fetter1,freilich10,middel_pra11,middel_pla}.

In the remainder of our work we will consider small clusters of
vortices $N=1,...,5$ with {\em same} charge. Without loss of
generality we consider $S_i=+1$ since the case $S_i=-1$ corresponds
to exactly the same dynamics if $t \rightarrow -t$.
It is straightforward to prove that the system of ODEs (\ref{ODEs}) possesses
two conserved quantities corresponding to the angular momentum $L$
and Hamiltonian $H$. The angular momentum assumes the form:
\begin{equation}
\label{L}
L = \sum_{i=1}^{N} r_i^2,
\end{equation}
and the Hamiltonian is given by:
\begin{eqnarray}
H&=&\frac{1}{2}\sum_{i=1}^{N} \ln(1-r_i^2) \nonumber \\
&-&\frac{c}{4}
\sum_{i =1}^N \sum_{j\ne i}
\ln(r_i^2+r_j^2-2 r_i r_j \cos(\theta_i-\theta_j)).
\label{H}
\end{eqnarray}
It is worth mentioning that
it is possible to reduce this Hamiltonian to $2N-2$ degrees of freedom
by using the conservation of angular momentum (\ref{L}) and introducing
it as a parameter and by defining
the relative angles $\delta_{ij}=\theta_{j}-\theta_{i}$
and thus effectively eliminating the polar angle of, let us say,
the first vortex by placing our dynamics in a frame rotating
with the first vortex.

From a mathematical viewpoint the parameter $c$ might be chosen
arbitrarily. However, all throughout our study we will use
the nominal value $c=0.1$ that has been shown to accurately
describe the experimental values for the quasi-2D case of rubidium atoms
under the experimental trapping
conditions of, e.g., Ref.~\cite{corot_prl}. It was also argued therein that
variations of the number of atoms of the condensate system would
only have a logarithmically weak effect on $c$, hence preserving this
constant value of $c$ provides a reasonable approximation.
For this fixed value of $c$ we
will vary the angular momentum between $0$ and $1$, i.e., we will use
the angular momentum as our {\it bifurcation parameter}.

At this point we should mention that it is straightforward to show that
the first of Eqs.~(\ref{ODEs}) can be written in vector form as follows:
\begin{equation}
\dot{r}^2_i=-c \left( \vec{r}_i \times  \sum_{j \neq i} \frac{\vec{r}_j}{\rho_{ij}^2} \right) \cdot \hat{e}_z,
\label{eq:vectr2}
\end{equation}
where $\hat{e}_z$ is the unit vector along $z$-direction. Equation~(\ref{eq:vectr2}) has a straightforward geometric
interpretation: each $r_i^2$ is conserved if the cross-product in the right-hand side of Eq.~(\ref{eq:vectr2})
vanishes; this is a necessary (but not sufficient)
condition for the existence of a fixed point.
There are two obvious cases for this cross product to become zero:
%
\begin{itemize}
\item The $\vec{r}_i$'s are all collinear.
\item The $\vec{r}_i$ define a regular polygon of order $N$ inscribed in a circle of radius $\sqrt{\frac{L}{N}}$.
\end{itemize}
%
Since the first case does {\it not} satisfy the second one of
Eqs.~(\ref{ODEs}) for general $N$, we
restrict our considerations to the second, more interesting candidate,
namely the polygonal case.
To establish its relevance,
let the center of the polygon be at the origin of the axes in $(x,y)$-plane,
and the considered polygon edge ($i$) lie at the positive $x$-axis. If $N$ is odd, all terms in the
sum $\sum_{j \neq i} \frac{\vec{r}_j}{r_{ji}^2}$ can be grouped into doublets (axially symmetric with
respect to the $x$-axis), such that their sum forms a vector parallel (or anti-parallel) to $\vec{r}_i$,
thus leading to the vanishing of the associated cross-product.
If $N$ is even, the grouping in doublets is possible for all $j$ except of one
which is anti-parallel to $\vec{r}_i$. Thus in either case ($N$ odd or even) the cross-product vanishes and, therefore,

the $r_i^2$ are conserved in this case.

Additionally, for small enough $N$, e.g., for $N=2$ or $N=3$,
if the vanishing of the cross product holds then the fixed-point equations for the polar angles $\theta_i$ are fulfilled as
well. In fact, in such cases, one can obtain analytical expressions for the fixed point configurations. However, we should
emphasize that these considerations only refer to the existence but {\it not} to the stability of the relevant symmetric
configurations (e.g., equilateral or isosceles triangles for $N=3$). For $N>3$ the equations resulting from the vanishing
of the cross product are less in number than the ones necessary to uniquely determine the fixed point configurations:
this is due to the fact that there exist $N$ equations for $\dot{r}_i=0$ and $N(N-1)/2$ for
$\dot{\theta}_i-\dot{\theta}_j=0$, hence it is not straightforward to
generalize this intriguing geometric interpretation beyond $N=3$.

Our ``deterministic''
computational approach in
seeking rigidly rotating states of the vortex particles
(effectively steady states in their relative angle variables) will be based
on the well-established continuation/bifurcation software AUTO~\cite{auto}.
We do not discuss AUTO further here, but direct the interested reader to
relevant resources mentioned above. Instead, we now provide more details
on our Monte-Carlo approach to identifying the system's ground state
as a function of $L$.

\section{The Monte-Carlo Method}
We employ the Metropolis Monte Carlo (MC) algorithm for obtaining the minimum energy state configurations of the vortices $\{r_i, \theta_i\}$ of the Hamiltonian (\ref{H})
for different numbers of vortices $N$. In order to exploit the conservation of the angular momentum, we introduce it as a parameter $L$ in the simulations and generate the $r_i$'s
through hyperspherical coordinates, thus enforcing the constraint $L=\sum_{i=1}^{N} r_i^2$. Due to the singularity of the Hamiltonian (\ref{H}) as
$r_i\rightarrow 1$, for our purposes we have had to restrict the values of $L$ in the interval $[0,1]$. It should be noted here that the idea of using MC type
approaches for particles interacting with logarithmic potentials
(and also sustaining an external confinement) stemmed from the pioneering
study of Ref.~\cite{peeters}, which, in turn, was motivated by experiments
(and phase transitions observed)
on systems of confined charged metallic balls~\cite{peeters11}.

In particular, we begin with setting four different initial configurations: the symmetric configuration, i.e.,
$\{ r_i=\sqrt{{L}/{N}},\theta_i={2(i-1) \pi}/{N}\}$ and three random ones. We then implement
the Metropolis algorithm at an (artificial)
ultra low temperature $kT=10^{-6}$, for each initial condition, until we
reach equilibrium, a fact that is checked by the convergence of the energy time series for the different random walks.
Each Monte Carlo step consists of the following procedural steps.

\medskip\noindent{\bf (1)}
We choose new configurations $\{ \tilde{r}_i, \tilde{\theta}_i\}$ such that they satisfy the constraints $L=\sum_{i=1}^{N} \tilde{r}_i^2$ and $0 \leq \tilde{r}_i \leq 1$.
A useful parametrization of the first condition in terms of the hyperspherical coordinates is defined through the relations:
\begin{eqnarray}
 \tilde{r}_1&=& \sqrt{L} \cos \phi_1 \nonumber \\
 \tilde{r}_2&=& \sqrt{L} \sin \phi_1 \cos \phi_2 \nonumber \\
 \tilde{r}_3&=& \sqrt{L} \sin \phi_1 \sin \phi_2 \cos \phi_3 \nonumber \\
&\cdots& \\
 \tilde{r}_{N-1}&=& \sqrt{L} \sin \phi_1  \sin \phi_2 \ldots \sin \phi_{N-2} \cos \phi_{N-1} \nonumber \\
 \nonumber
 \tilde{r}_N &=& \sqrt{L} \sin \phi_1  \sin \phi_2 \ldots \sin \phi_{N-2} \sin \phi_{N-1}.
\end{eqnarray}
Note that we arrive at $N-1$ independent angles and thus $\tilde{r}_N$ is completely determined by the knowledge of the other $\tilde{r}_i$'s. In the following we denote the prefactors of
 $\cos \phi_i$
with $\alpha_i$, i.e., $\tilde{r}_i= \alpha_i \cos \phi_i$.
We have to also fulfill the second constraint.
It is easily shown that the requirement $\tilde{r}_i \geq 0$ is fulfilled, without loss of generality, by constraining the angles $\phi_i$ to the first quadrant, thus
if $0 \leq \phi_i \leq \frac{\pi}{2}$.
In order to satisfy the condition  $\tilde{r_i} \leq 1$ we need:
\[\alpha_i \cos \phi_i \leq 1 \quad\Rightarrow\quad \cos \phi_i \leq \frac{1}{\alpha_i}.\]
However, since the $\tilde{r}_i$ are determined recursively we also need to ensure that with a random choice of $\tilde{r}_j$ all the  $\tilde{r}_i$'s can be less than $1$.
This is not trivial especially for large values of $L$.
Beginning with $\tilde{r}_1$ we have that
\begin{eqnarray}
\sum_{i=2}^{N} \tilde{r}_i^2 \leq N-1
&\Rightarrow& L-\tilde{r}_1^2 \leq N-1
\nonumber
\\[-1.0ex]
\nonumber
&\Rightarrow& \tilde{r}_1^2 \geq \alpha_1^2-(N-1).
\end{eqnarray}
Similarly, for $\tilde{r}_2$:
\begin{eqnarray}
\sum_{i=3}^{N} \tilde{r}_i^2 \leq N-2
&\Rightarrow& L-\tilde{r}_1^2 -\tilde{r}_2^2\leq N-2
\nonumber
\\[-1.0ex]
\nonumber
&\Rightarrow&  L \sin^2 \phi_1 -\tilde{r}_2^2\leq N-2
\nonumber
\\[1.0ex]
\nonumber
&\Rightarrow& \tilde{r}_2^2 \geq \alpha_2^2-(N-2).
\end{eqnarray}
Recursively, this leads to:
\[
\tilde{r}_i^2 \geq \alpha_i^2-(N-i) \quad\Rightarrow\quad \cos \phi_i \geq \sqrt{\frac{\alpha_i^2 -(N-i)}{\alpha_i^2}}.
\]
Gathering all these conditions together we are led to the requirement:
\[
M_i \leq \cos \phi_i \leq m_i,
\]
where
\[
M_i \equiv \sqrt{\max\left\{0,{\frac{\alpha_i^2 -(N-i)}{\alpha_i^2}}\right\}}\,,
\]
and
\[
m_i \equiv \min\left\{1,\frac{1}{\alpha_i}\right\}.
\]
We thus generate the $\left\{ \tilde{r}_i\right\}$ in ascending order beginning with $\tilde{r}_1$, by choosing $\phi_i$'s randomly from a uniform distribution subject to the condition:
\begin{equation}
\cos^{-1} \left(m_i \right) \leq \phi_i \leq \cos^{-1} \left(M_i \right).
 \end{equation}

Concerning the angles $\tilde{\theta}_i$'s, for an 
index $j$ we choose $\tilde{\theta}_j$ randomly from a uniform distribution
$0 \leq \tilde{\theta}_j \leq 2 \pi$. For all the other angles the
old values are kept, namely $\tilde{\theta}_i = \theta_i$ for $ i \neq j$.

\medskip\noindent{\bf (2)}
We then calculate the difference $\Delta E =E_{\rm new}- E_{\rm old}$,
where $E_{\rm old}=H\left(  \left\{r_i, \theta_i\right\} \right)$
is the energy of the old configuration and
$E_{\rm new}=H\left(  \{\tilde{r}_i, \tilde{\theta}_i \} \right)$
is the energy of the new one.

\medskip\noindent{\bf (3)}
If $\Delta E \leq 0$ the new configuration is accepted, i.e. $r_i=\tilde{r}_i$, $\theta_i=\tilde{\theta}_i$. Otherwise, we accept the new configuration with
a probability $P$ given by the Boltzmann factor $P=\exp (-\beta E)$,
where $\beta=1/kT$.

\medskip

After reaching equilibrium, in our case typically after $5 \cdot 10^6$ MC
steps, we have practically the configurations for $T \approx 0$,
i.e., the minimum energy configurations sought. In order to optimize our results in this step we
perform a final MC simulation at $T=0$. This deterministic local search reduces
some of the fluctuations and allows us to obtain the
minimum configuration with a desirable accuracy, which in our simulations leads to an error of
order $10^{-4}$.

We remark that the MC algorithm always converges to a minimum but it does not distinguish
between local and global ones. In order to handle this problem usually a large number of initial conditions,
or the use of more sophisticated techniques like simulated annealing are required. However,
for the cases examined here with a  small number of particles, four initial conditions are proven to be sufficient
for identifying the global minimum. 
This is also justified by the coincidence of the MC results with those 
obtained by the solutions of the corresponding ODEs presented in the following section.

\begin{figure}[htbp]
\begin{center}
\includegraphics[width=8cm]{\rootfig 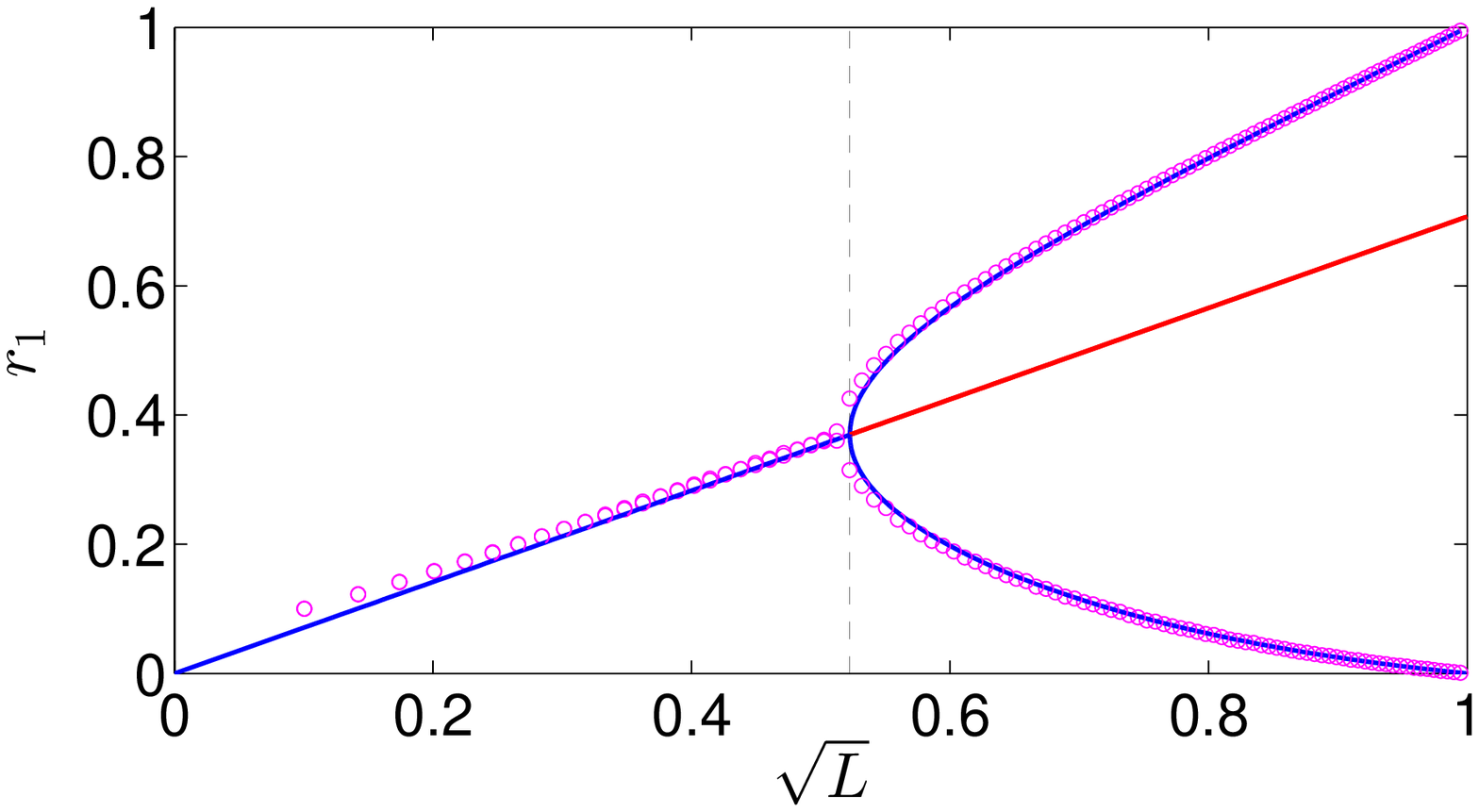}\\[1.0ex]
\includegraphics[width=8cm]{\rootfig 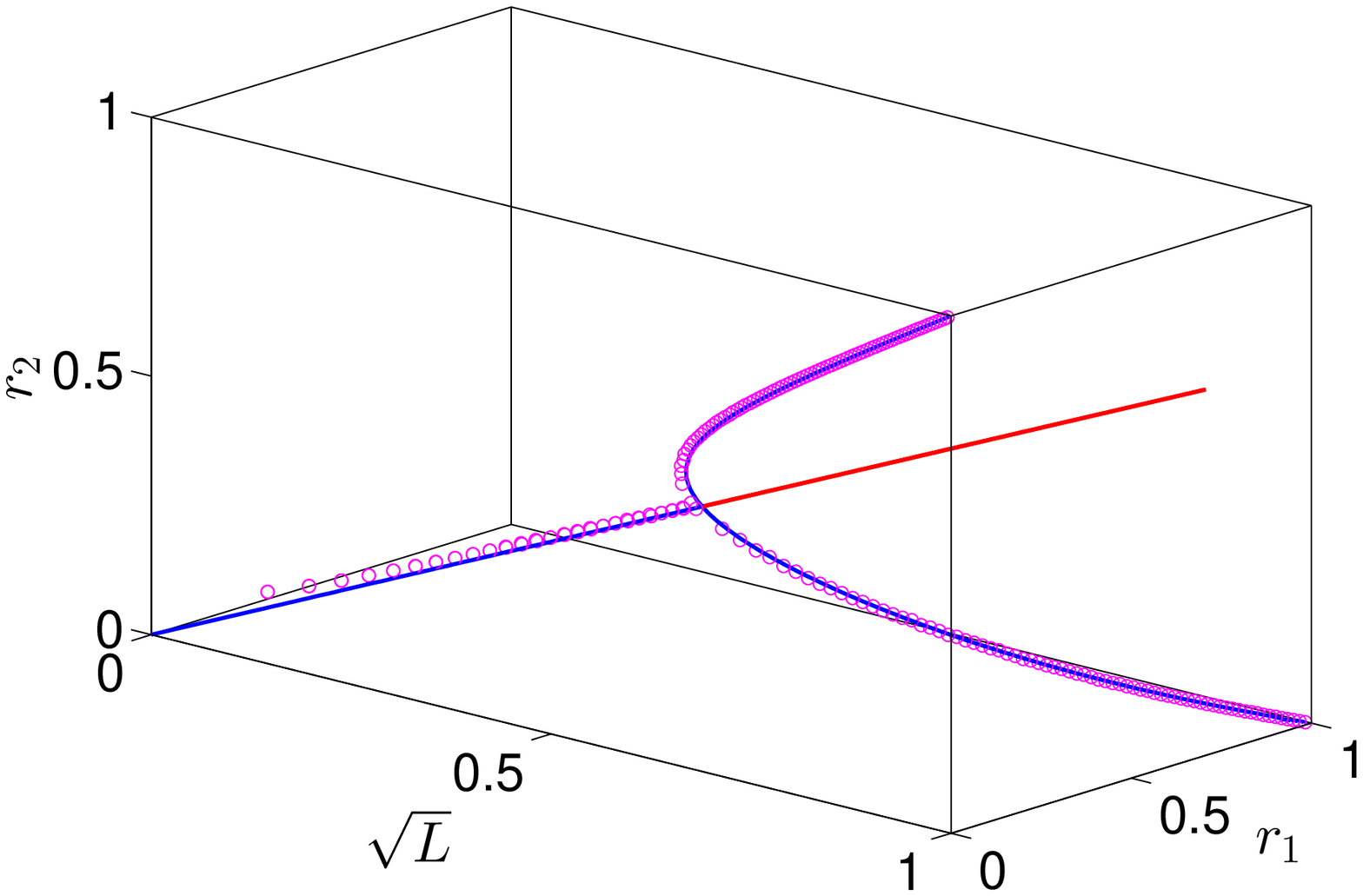}
\end{center}
\caption{(Color online).
Bifurcation scenario for $N=2$ same charge vortices and
corresponding MC simulations. The bifurcation diagram,
as a function of the square root of the angular momentum of the
system ($\sqrt{L}$) is increased,
obtained from the corresponding ODEs (\ref{ODEs})
is depicted by the solid lines where blue denotes a
stable branch and red an unstable branch.
The MC simulation results are depicted with the
small magenta circles.
The critical point beyond which (through a
supercritical pitchfork bifurcation) the asymmetric configurations
arise --and become a ground state of the system--
is indicated
by the vertical dashed line.
Note how the MC simulations nicely follow the stable branches of the
bifurcation diagram.
}
\label{N2}
\end{figure}

\begin{figure*}[htbp]
\begin{center}
\includegraphics[width=8cm]{\rootfig 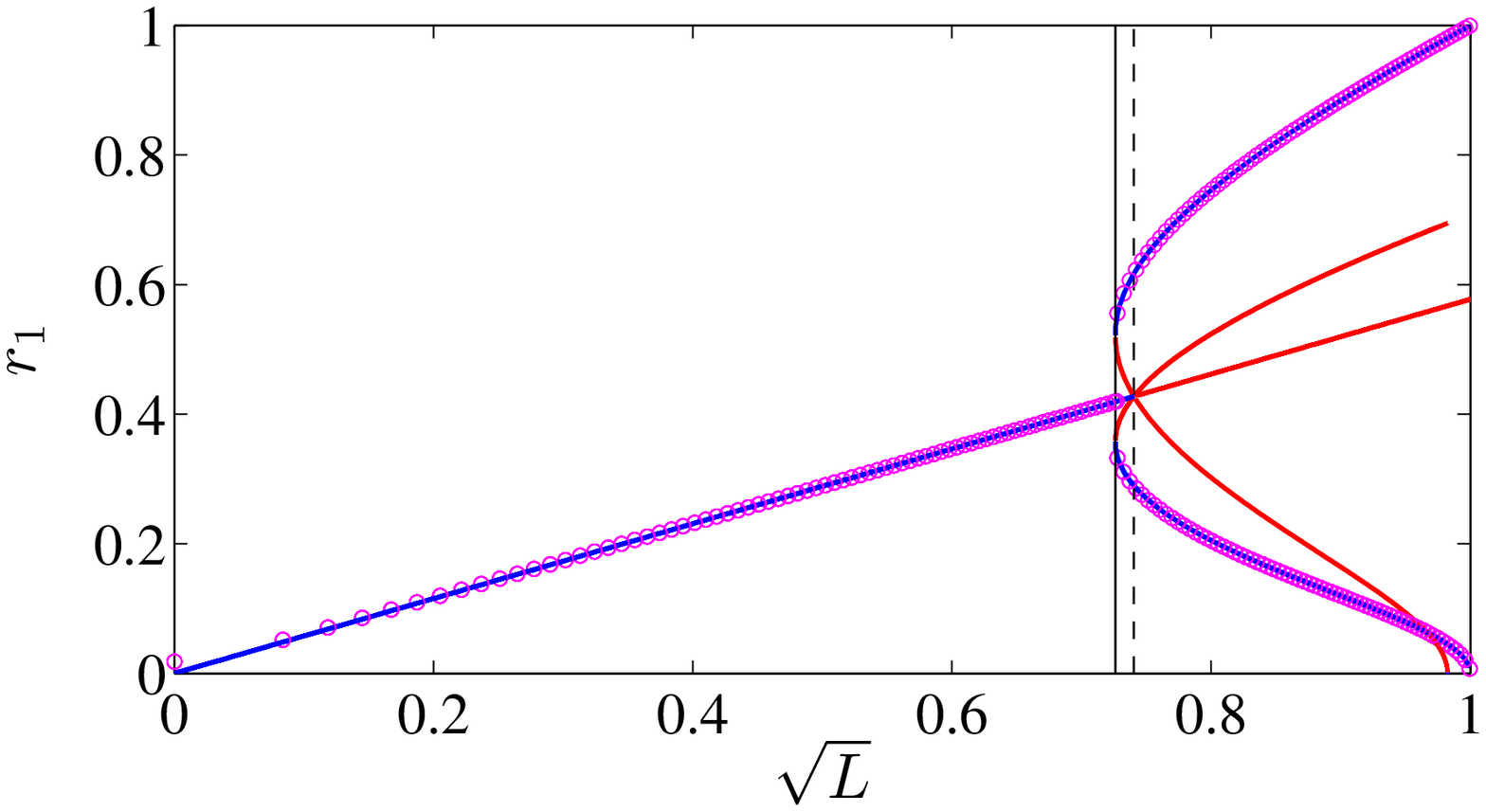}~~
\includegraphics[width=8cm]{\rootfig 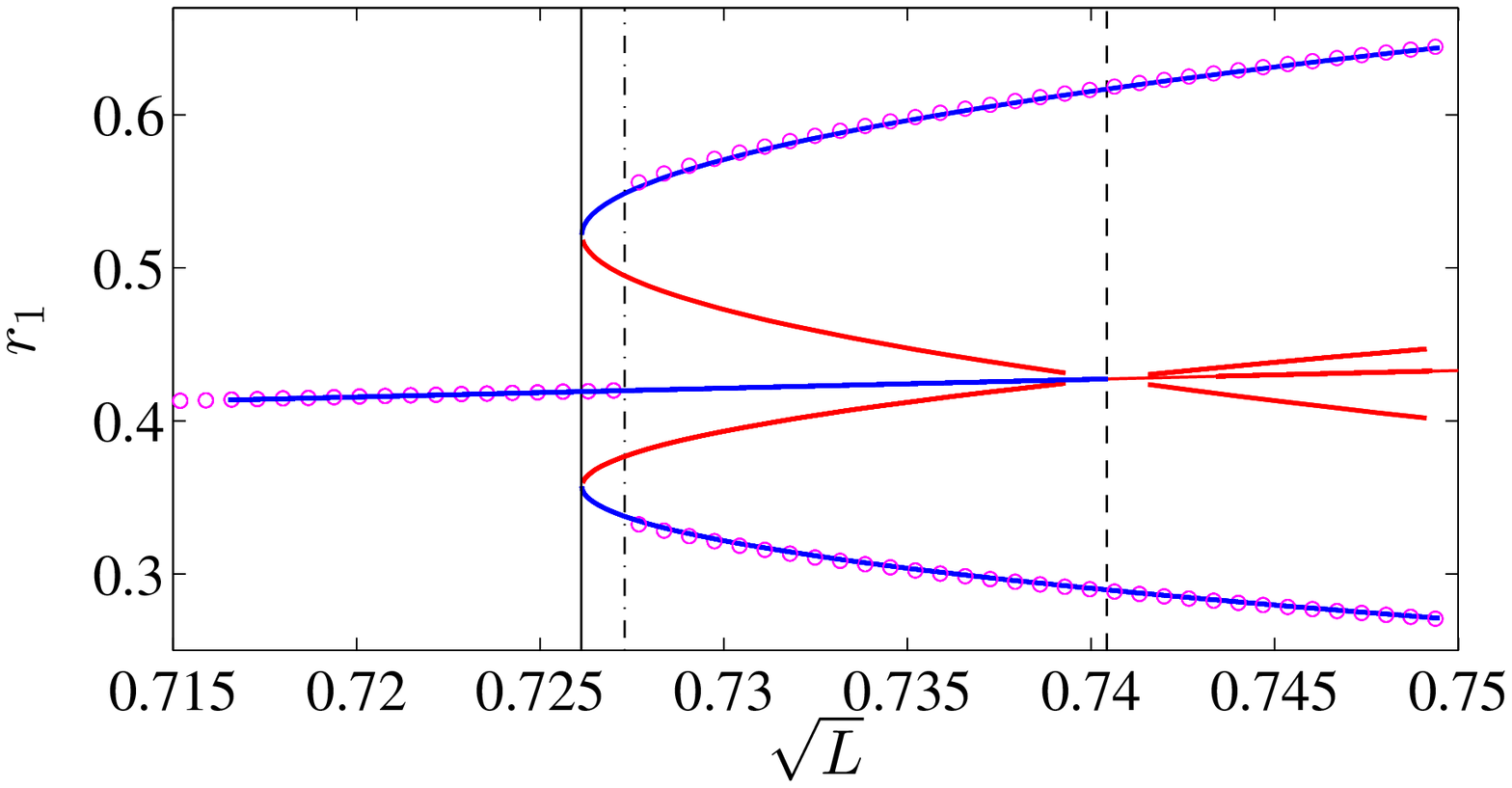}\\[1.0ex]
\includegraphics[width=8cm]{\rootfig 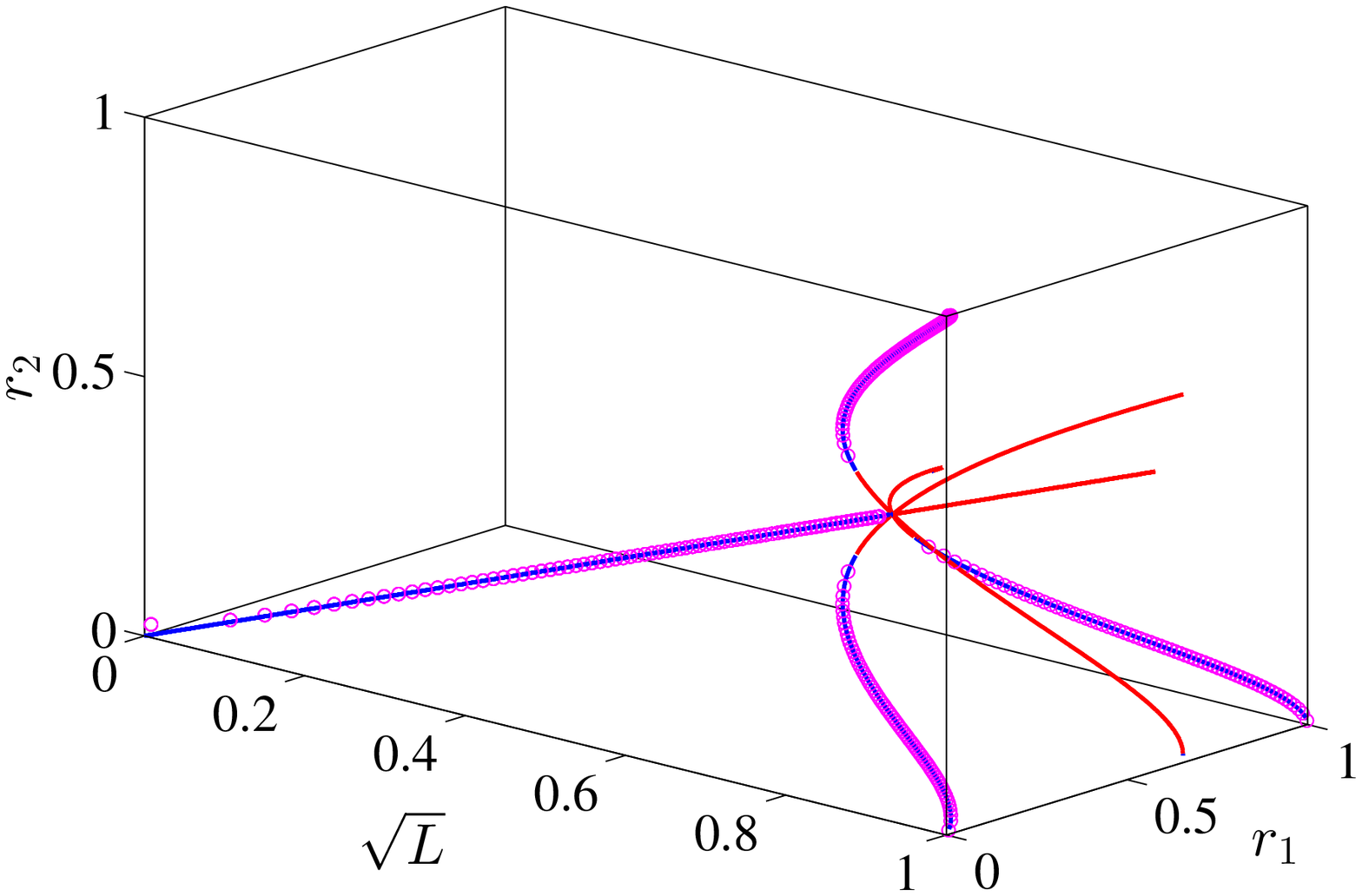}~~
\includegraphics[width=8cm]{\rootfig 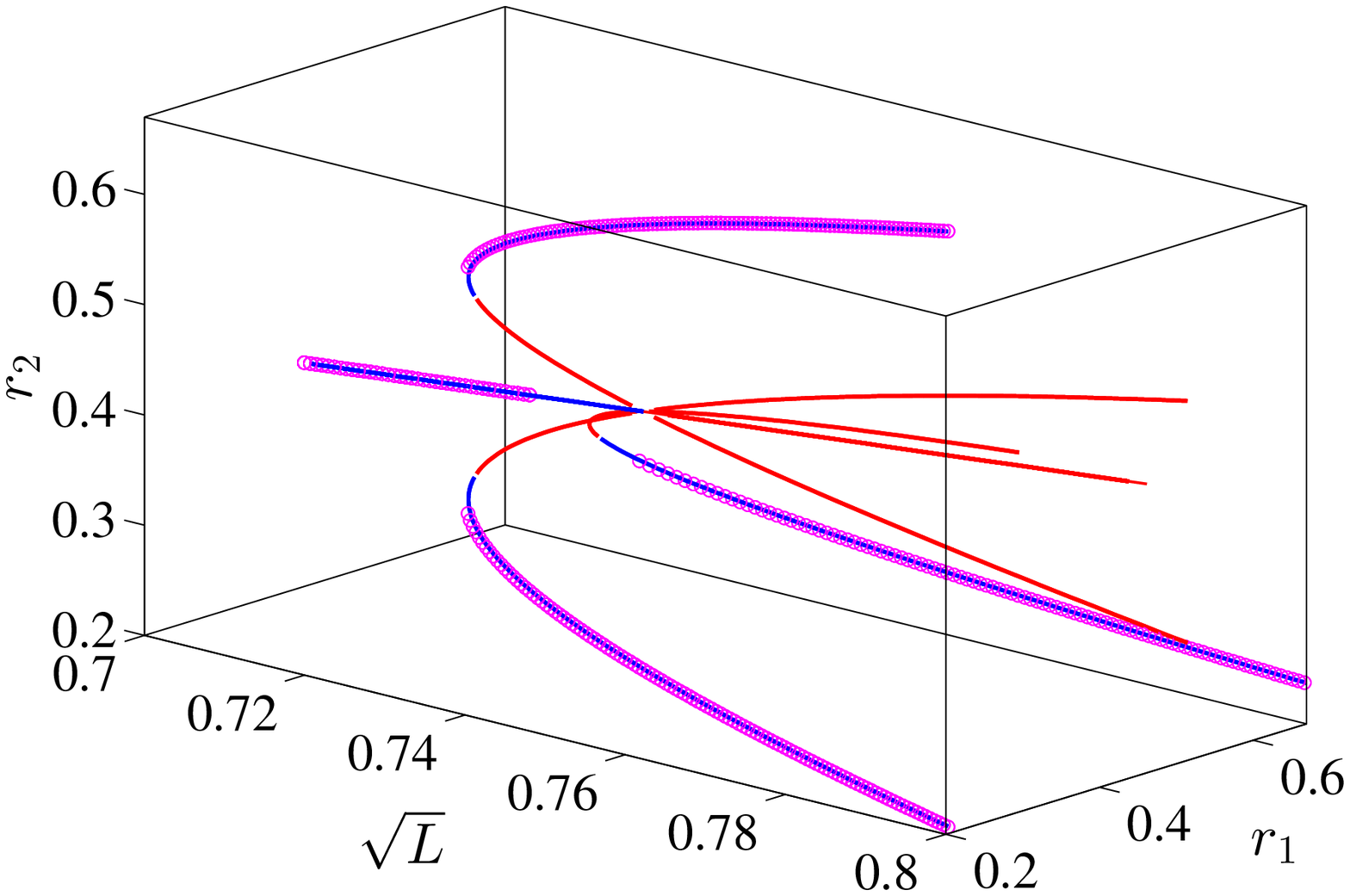}
\end{center}
\caption{(Color online).
Bifurcation scenario for $N=3$ same charge vortices and
corresponding MC simulations. Same notation as in Fig.~\ref{N2}.
The left panels correspond to the entire domain $0\leq L \leq1$
while the right panels depict a zoomed in version near the
bifurcation region. The top panels show the bifurcation
diagrams in a planar view, namely $r_i$ vs.~$\sqrt{L}$.
The bottom panels depict the bifurcation diagram in
the three dimensional space ($\sqrt{L},r_1,r_2$)
where the the $C_3$ symmetry of the solution is
clearly visible.
The thin dashed vertical line corresponds to the
symmetry breaking bifurcation where the symmetric (equilateral
triangle) loses its stability.
The thin solid vertical line depicts
the location of the saddle-center bifurcation of
collisions between asymmetric (isosceles triangle) solutions.
The thin dash-dotted vertical line corresponds to
the location where the MC simulation switches branches
(i.e., location where the energy minimum configuration switches from
an equilateral to an isosceles one).
}
\label{N3}
\end{figure*}

\section{Results}
We now present our results in terms of the (numerically) exact
bifurcation diagrams of the coupled systems of ODEs (\ref{ODEs}) and
the corresponding approximate (ground state) phase diagrams obtained by the
MC methodology described in the previous section. We will perform
this comparison for $N=2,3$, and $4$ vortices and present the
MC results for $N=5$ vortices. In all of these cases, we complement
our computations with analytical results, wherever possible.

\subsection{The N=2 Vortex Case}

We start by examining for completeness (and in order to set the stage for
follow-up observations) the case of $N=2$ that was
previously considered in some detail in
Ref.~\cite{corot_prl}. As discussed in that
work, for values of the angular momentum $L < L_{cr}^{(2)}
\equiv  2 \sqrt{c}/(\sqrt{c} + 2)$,
i.e., for radial displacements of the vortices $r < r_{cr}^{(2)}
\equiv \sqrt{\sqrt{c}/(\sqrt{c} + 2)}$,
the {\rm symmetric} rigidly rotating vortex state,
namely two vortices at equal distances from the center of the trap,
is {\em stable}.
However, for radii (or angular momenta) above this critical
point, the symmetric state becomes {\it structurally unstable}
and gives rise, through a super-critical, for the range
of $c$'s of relevance to the experiment,
pitchfork bifurcation (i.e., a spontaneous
symmetry breaking), to the emergence of {\em asymmetric}, yet
still anti-diametric, rigidly rotating
states. In the latter, one of the vortices is always further away
from the origin, say $r_1$, while the other is closer,
say $r_2$ ($r_2<r_1$), such that the angular momentum constraint
$L= r_1^2+ r_2^2$ is satisfied and also
\begin{eqnarray}
-r_1 r_2 (r_1+r_2)^2 + c (1-r_1^2) (1-r_2^2) =0.
\label{eqn2}
\end{eqnarray}
This analytical expression identified in Ref.~\cite{corot_prl} by means
of a direct solution of the equations of motion, along with the
angular momentum constraint and the necessity that $\delta_{12}= \pi$
(i.e., anti-diametric vortices) fully characterize the class of
asymmetric solutions in the case of $N=2$.

Our Monte-Carlo analysis does an excellent job of capturing the
relevant minimizers of the energy. As it is clear from the (magenta) data
points of Fig.~\ref{N2}, up to the critical point $L_{cr}^{(2)}$ (see
vertical dashed line),
the Monte-Carlo computation follows the symmetric branch
(see blue line for $L<L_{cr}^{(2)}$),
while past the critical point, it follows the
newly emergent and stable (see blue curves for $L>L_{cr}^{(2)}$) asymmetric
branch arising from the pitchfork bifurcation.
This case serves as a useful benchmark between the analysis
and the numerical MC computation, and as a prototypical example
of the phenomenology that will follow, involving the
spontaneous emergence of asymmetric rigidly rotating
states and which will be progressively
more complex as the number of vortices $N$ increases.

\subsection{The N=3 Vortex Case}

\begin{figure}[htbp]
\begin{center}
\includegraphics[width=8cm]{\rootfig 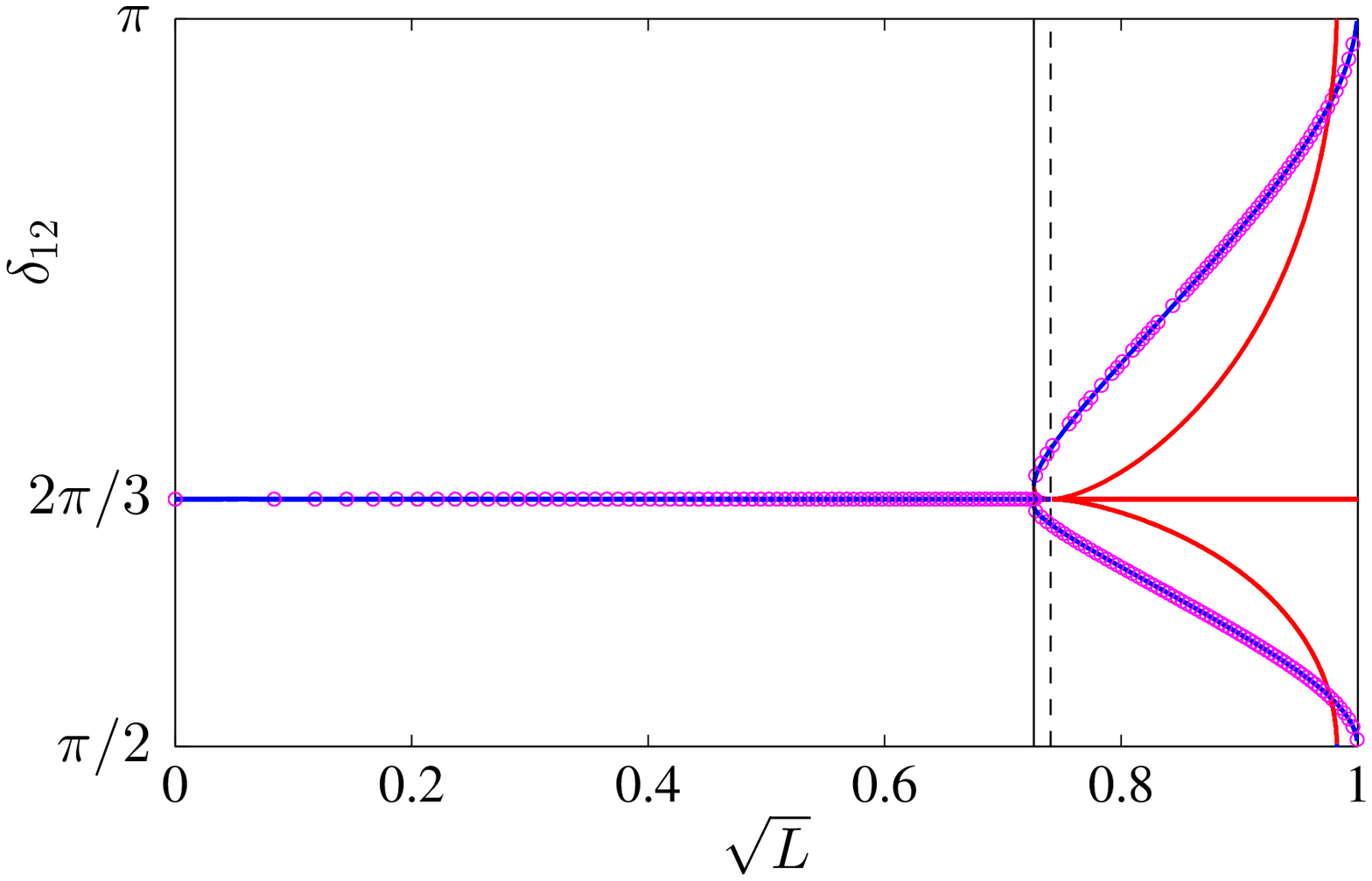}
\includegraphics[width=8cm]{\rootfig 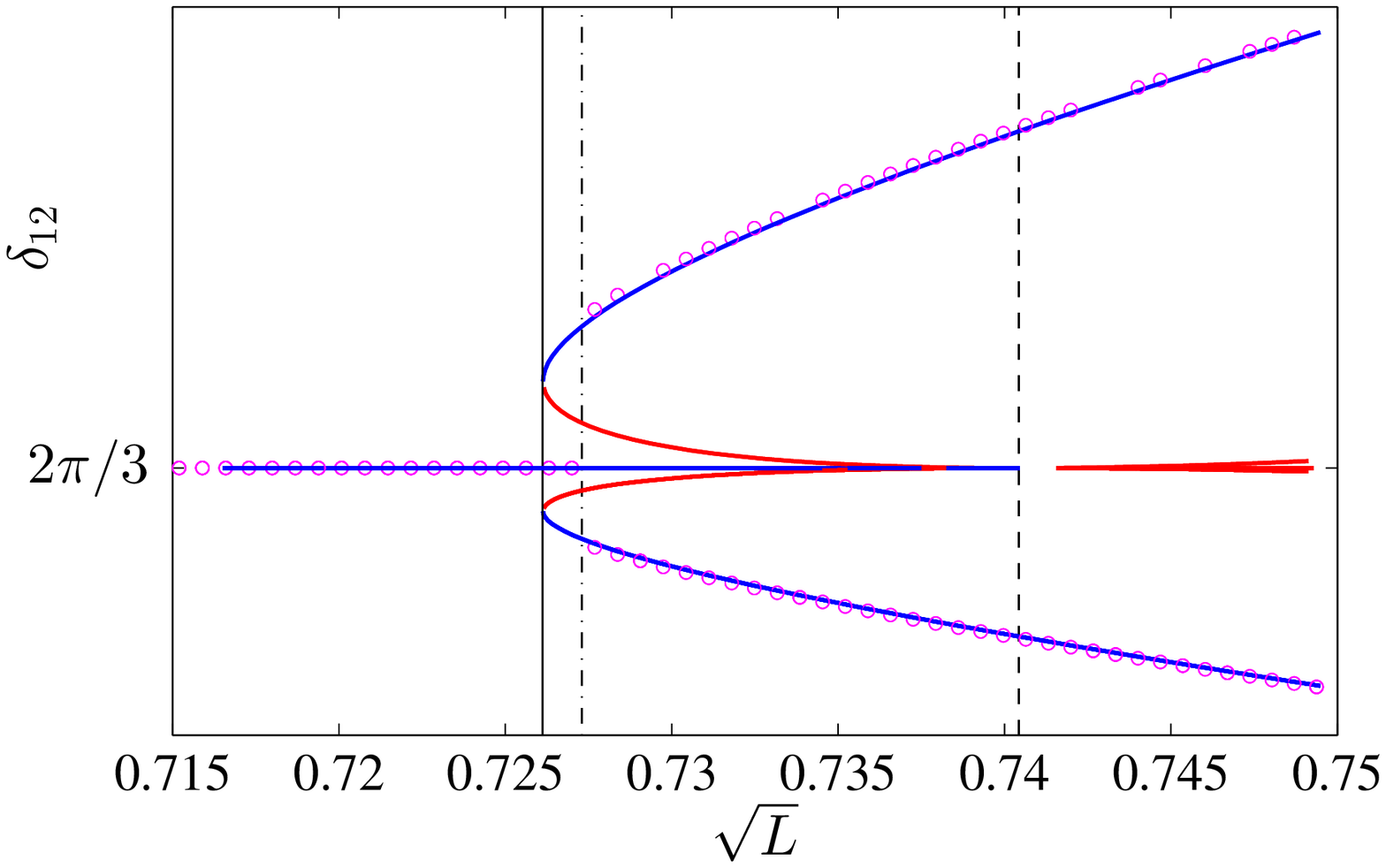}
\end{center}
\caption{(Color online).
Bifurcation scenario for the relative angles between vortices $\delta_{12}$
as a function of the (square root) of the angular momentum
for $N=3$ vortices.
The data is the same as Fig.~\ref{N3} and the notation is also the same.
The top panel is the full view while the bottom panel depicts a zoomed in
version around the bifurcation region.
}
\label{N3pi}
\end{figure}

For $N=3$, the symmetric rigidly rotating
solution naturally persists (in fact, it persists
for all the $N$'s that we have considered) with the relevant
inter-vortex angle being $\delta_{ij}=2 \pi/N = 2 \pi/3$ in this case.
The angular momentum constraint for this equilateral solution
reads $L=N r^2= 3 r^2$.
The stability of this solution can be also identified analytically.
In particular, there is an eigenfrequency associated with it
assuming the analytical form:
\begin{eqnarray}
\omega^2=\frac{c^2}{r^4} - \frac{2 c}{(1-r^2)^2}.
\label{eqn30}
\end{eqnarray}
The zero crossing of this squared eigenfrequency at
$r= r_{cr}^{(3)} \equiv \sqrt{{\sqrt{c}}/({\sqrt{c} + \sqrt{2}})} = 0.4275$
yields the destabilization point of the equilateral triangle.
The corresponding critical angular momentum satisfies:
$\sqrt{L_{cr,1}^{(3)}} = \sqrt{3} r_{cr}^{(3)} =  0.7404$.
This critical threshold is depicted by the thin vertical
dashed line in Fig.~\ref{N3} (and also Fig.~\ref{N3pi}).


\begin{figure*}[htbp]
\begin{center}
\includegraphics[width=8cm]{\rootfig 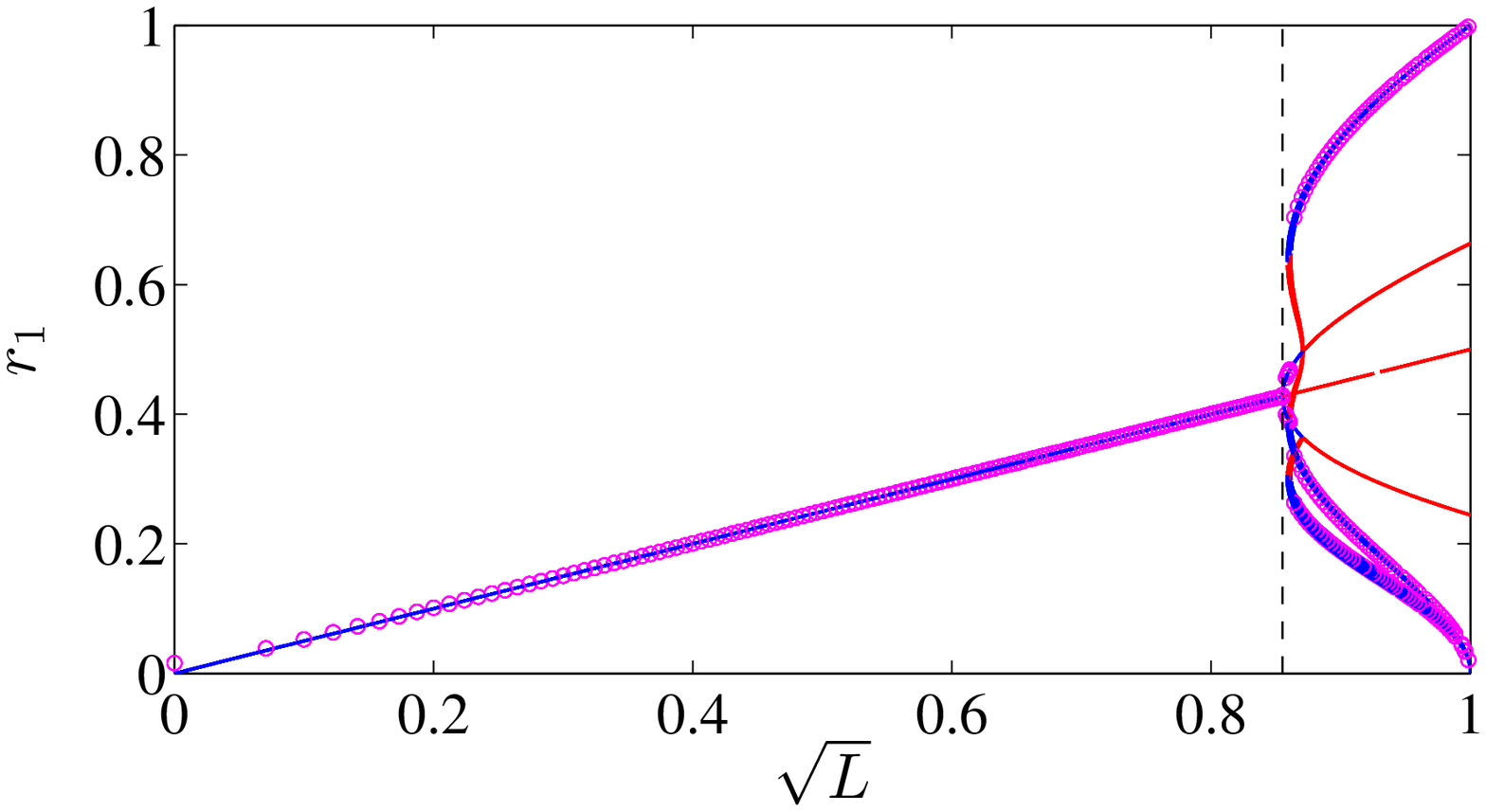}~~
\includegraphics[width=8cm]{\rootfig 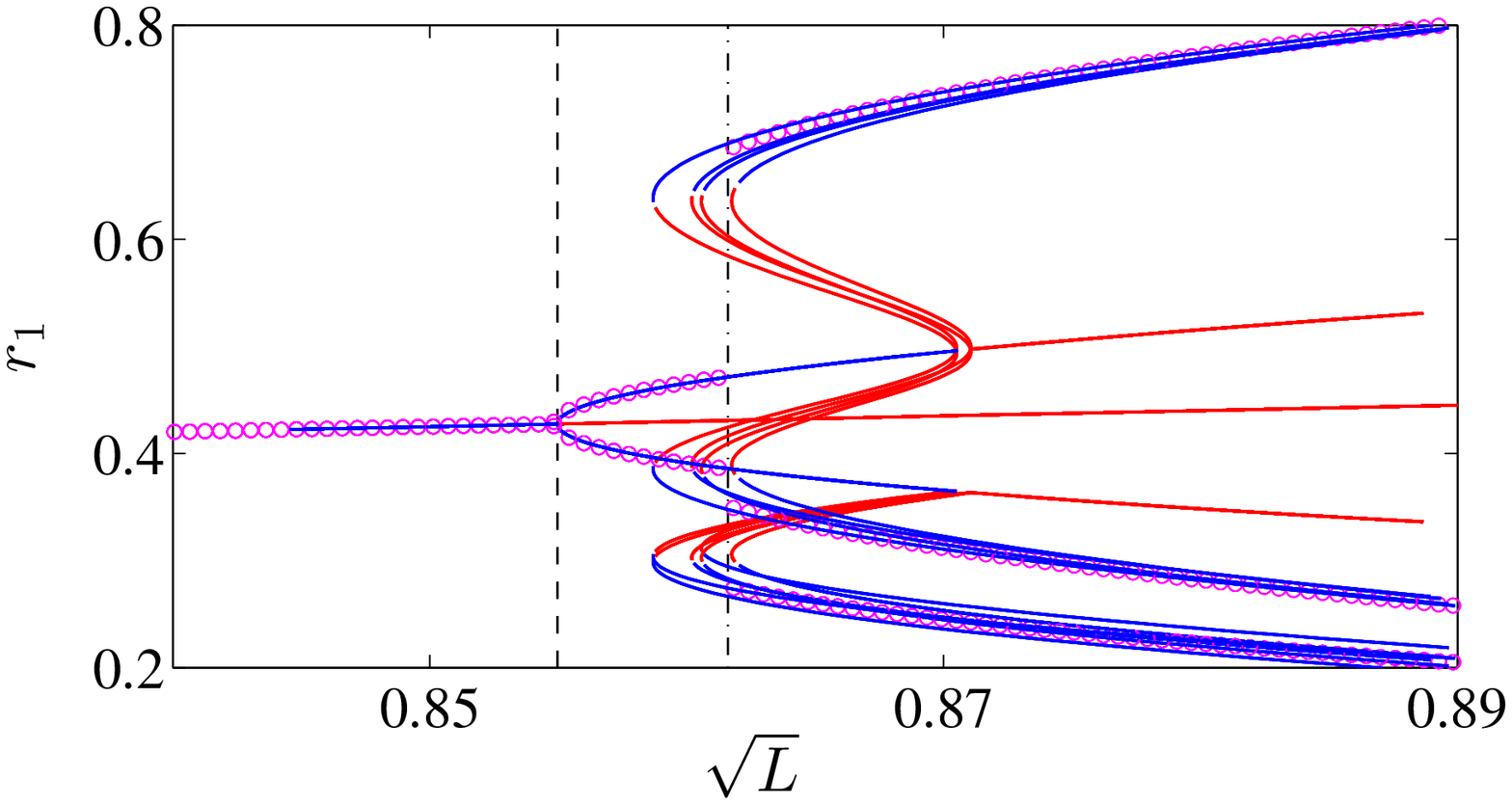}\\[1.0ex]
\includegraphics[width=8cm]{\rootfig 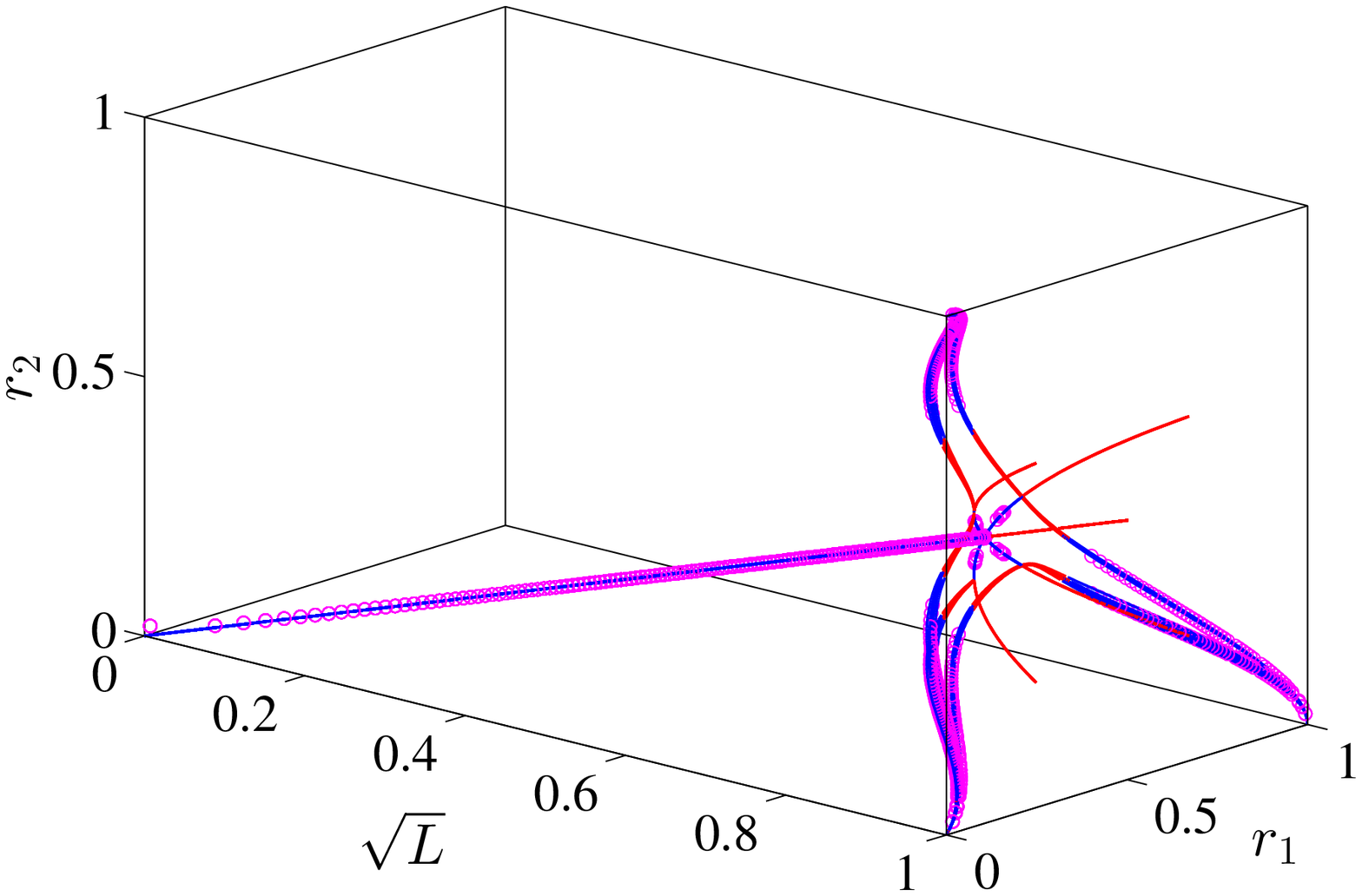}~~
\includegraphics[width=8cm]{\rootfig 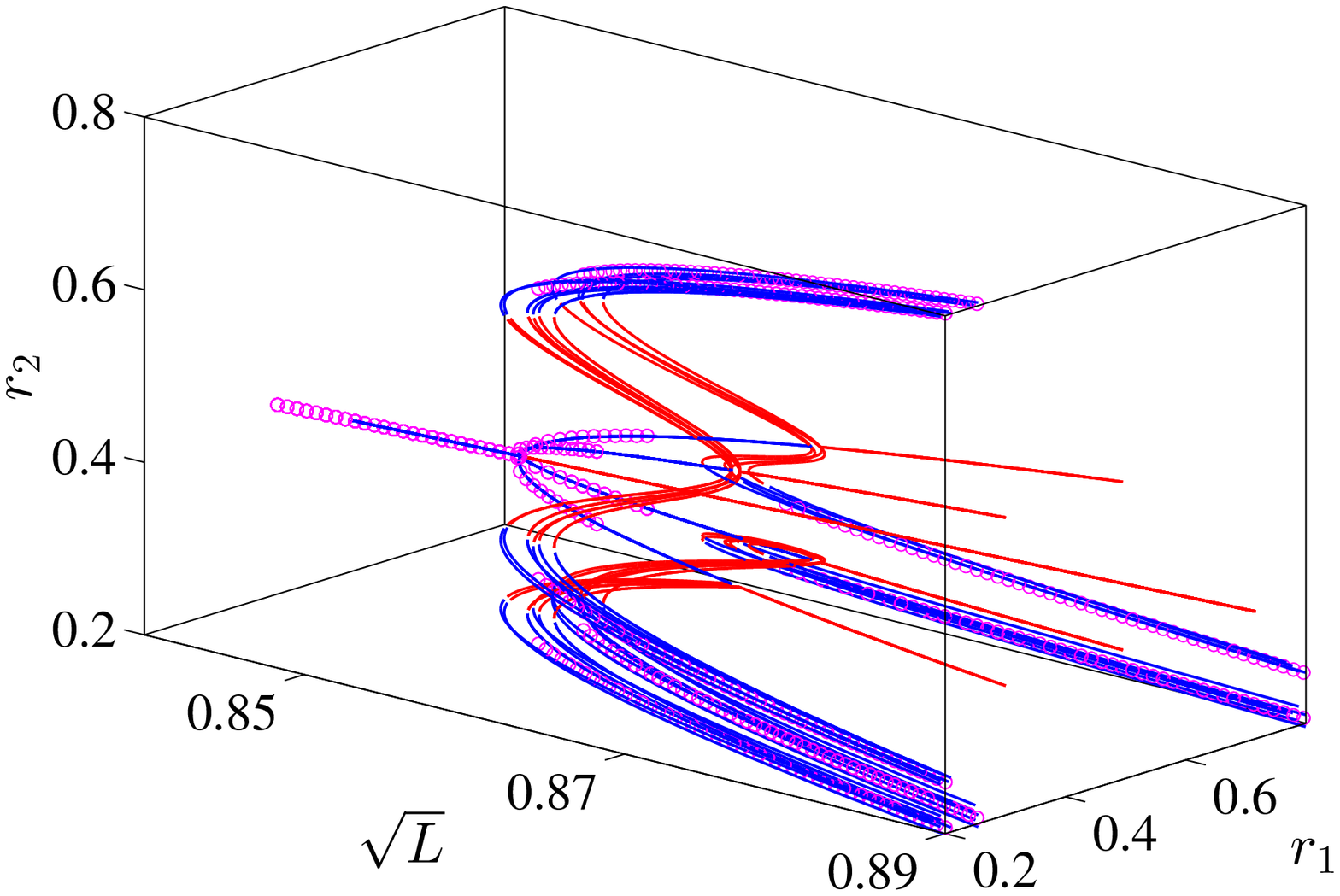}
\end{center}
\caption{(Color online).
Bifurcation scenario and
corresponding MC simulations showing the complete dynamical
picture associated with the transitions/instabilities in the case of
$N=4$. Similarly to the $N=3$ case, the top panels show a planar representation
of the solutions using only $r_1$ as a function of $\sqrt{L}$, while
the bottom panels relay a 3-dimensional variant thereof with $r_2$,
as a function of $r_1$ and $\sqrt{L}$.
The blue lines are stable branches, while the red lines
represent the unstable branches and the Monte-Carlo data are overlayed
using small magenta circles.
This conveys, not only how new branches (such as the rhombus and
general quadrilateral) emerge through suitable bifurcations (supercritical
pitchfork or saddle-center, respectively), but also when they become
the global energy minimizers and hence are followed by the Monte-Carlo
simulation.
Specifically, the thin dashed vertical line corresponds to the
symmetry breaking bifurcation where the symmetric (square) configuration
loses its stability towards the newly created rhombus configuration,
while the thin dash-dotted vertical line corresponds to
the location where the MC simulation switches branches
(i.e., transitions from the rhombic configuration
to a more general quadrilateral without any apparent symmetry).
}
\label{N4}
\end{figure*}

The $N=3$ case is richer than 
$N=2$. In particular, in addition to the
symmetry-breaking pitchfork bifurcation that destabilizes the symmetric
(equilateral triangle) solutions, which is equivalent to the
one we described for the $N=2$ case, there is another bifurcation.
This secondary bifurcation happens at the critical point $L_{cr,2}^{(3)}$
where new solutions emerge
(see threshold depicted by the thin solid vertical line in Fig.~\ref{N3}).
In fact, this is a pair of solutions
with a trilateral symmetry ($C_3$) corresponding to the three
possible {\it isosceles triangles} of vortices that can
emerge as rigidly rotating solutions in the system. For
these solutions, one of the vortices is at, say a longer
distance from the origin, $r_1$, while the other two
are at, say a shorter distance $r_2$. Then the angular
momentum constraint reads $L=r_1^2 + 2 r_{2}^2$, while
the following conditions completely specify the relevant
solution in an analytical form:
\begin{eqnarray}
\delta_{12}  &=& \delta_{31} = \frac{d \pm \sqrt{d^2+8}}{4},
\label{eqn31}
\notag
\\[2.0ex]
\delta_{23}  &=& 2 \pi - 2 \delta_{12},
\label{eqn32}
\\[2.0ex]
\notag
c &=& \frac{3 r_1^4 r_2^2 +3 r_1^2 r_2^4 -
a\, r_1^3 r_2^3
}{(1-r_1^2) (1-r_2^2)
(r_1^2 - r_2^2 - a\, r_1 r_2)
}
\label{eqn33}
\end{eqnarray}
where $\cos (d)\equiv (r_1^2+r_2^2)/(2 r_1 r_2)$,
and $a^2\equiv {r_1^4 + 34 r_1^2 r_2^2 + r_2^4}$.

While we have attempted to identify this secondary critical
point in a tractable analytical form,
it has not been possible given the complexity of the above solution.
Nevertheless, we have been able to identify numerically that the
relevant bifurcation that leads to the emergence of the isosceles
triangles is a saddle-center one. Namely, each of the 3
(rotated by 120$^{\,0}\deg$) triangles comes with an ``unstable partner''.
This saddle-center bifurcation
arises numerically at $L_{cr,2}^{(3)}=0.527 \Rightarrow
\sqrt{L_{cr,2}^{(3)}}=0.726$, as illustrated by the thin solid
vertical line in Fig.~\ref{N3}.
It is in fact very close to this point that the Monte-Carlo
computation will jump at this newly arising (stable such)
branch. That is to say, almost as soon as the branch is born, it becomes
the global minimum of the energy surface.
%
%
Remarkably, it is the
unstable partner of these isosceles saddle-center pairs
which collides with the symmetric, equilateral solution
at $L_{cr,1}^{(3)} = 0.548 \Rightarrow \sqrt{L_{cr,1}^{(3)}}=0.7404$
(see threshold depicted by a thin dashed vertical line in Fig.~\ref{N3}).
Our dynamical and eigenvalue computations of Fig.~\ref{N3}
capture this transition but the Monte-Carlo is entirely insensitive
to this step.  This, in turn, suggests the relevance of the Monte-Carlo
as a convenient tool for identifying the global energy minimum of the system
but also the usefulness of the full dynamical systems analysis provided
herein as a means of identifying metastable states and transitions
between them. The combination of the two unveils some of the complexities
of the full energy surface. While Fig.~\ref{N3} focuses on the
dependence of the radii of the particles as a function of
the angular momentum $L$, Fig.~\ref{N3pi} shows the corresponding
relative angles between vortices ($\delta_{ij}$). These
deviate from their equilateral value of $2 \pi/3$ in
an asymmetric manner, revealing the isosceles character of the
triangle given that out of the three equal angles, only two remain
equal while the third acquires a different value.

\subsection{The N=4 Vortex Case}

We now turn to the more complex case of $N=4$. Here, too, the
symmetric solution exists with $L = 4 r^2$ and $\delta_{ij}= \pi/2$.
However, the linearization around it now features two internal
modes. The first of them has the frequency
\begin{eqnarray}
\omega_1^2=\frac{2 c^2}{r^4} - \frac{4 c}{(1-r^2)^2},
\label{eqn40}
\end{eqnarray}
and remarkably crosses zero (and thus marks the
critical point for the destabilization of the configuration)
at the {\em same}
point as the $N=3$ case, i.e., at
$r= r_{cr,1}^{(4)} \equiv \sqrt{{\sqrt{c}}/(\sqrt{c} + \sqrt{2})}$,
which, however, now corresponds to the higher angular momentum
$L_{cr,1}^{(4)}= 4 {\sqrt{c}}/({\sqrt{c} + \sqrt{2}})$. The second
of these critical points corresponds to the eigenfrequency:
\begin{eqnarray}
\omega_2^2=\frac{9 c^2}{4 r^4} - \frac{3 c}{(1-r^2)^2}
\label{eqn41}
\end{eqnarray}
which vanishes at $r^2= (r_{cr,2}^{(4)})^2 \equiv {\sqrt{3 c}}/({\sqrt{3 c} + 2})$;
this second critical point does not appear to be of particular
interest to our study here.

\begin{figure}[htbp]
\begin{center}
\includegraphics[width=8cm]{\rootfig 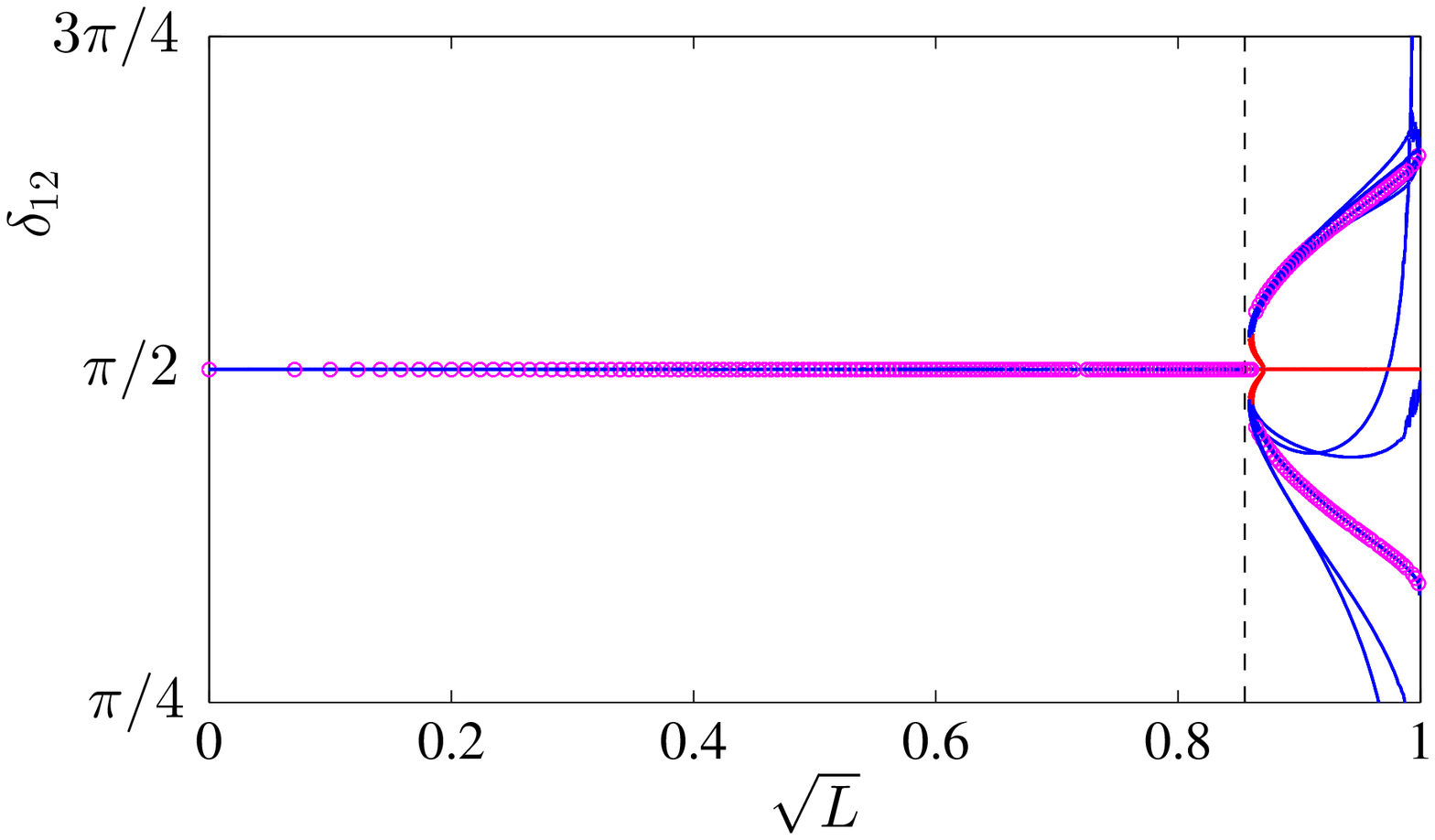}
\includegraphics[width=8cm]{\rootfig 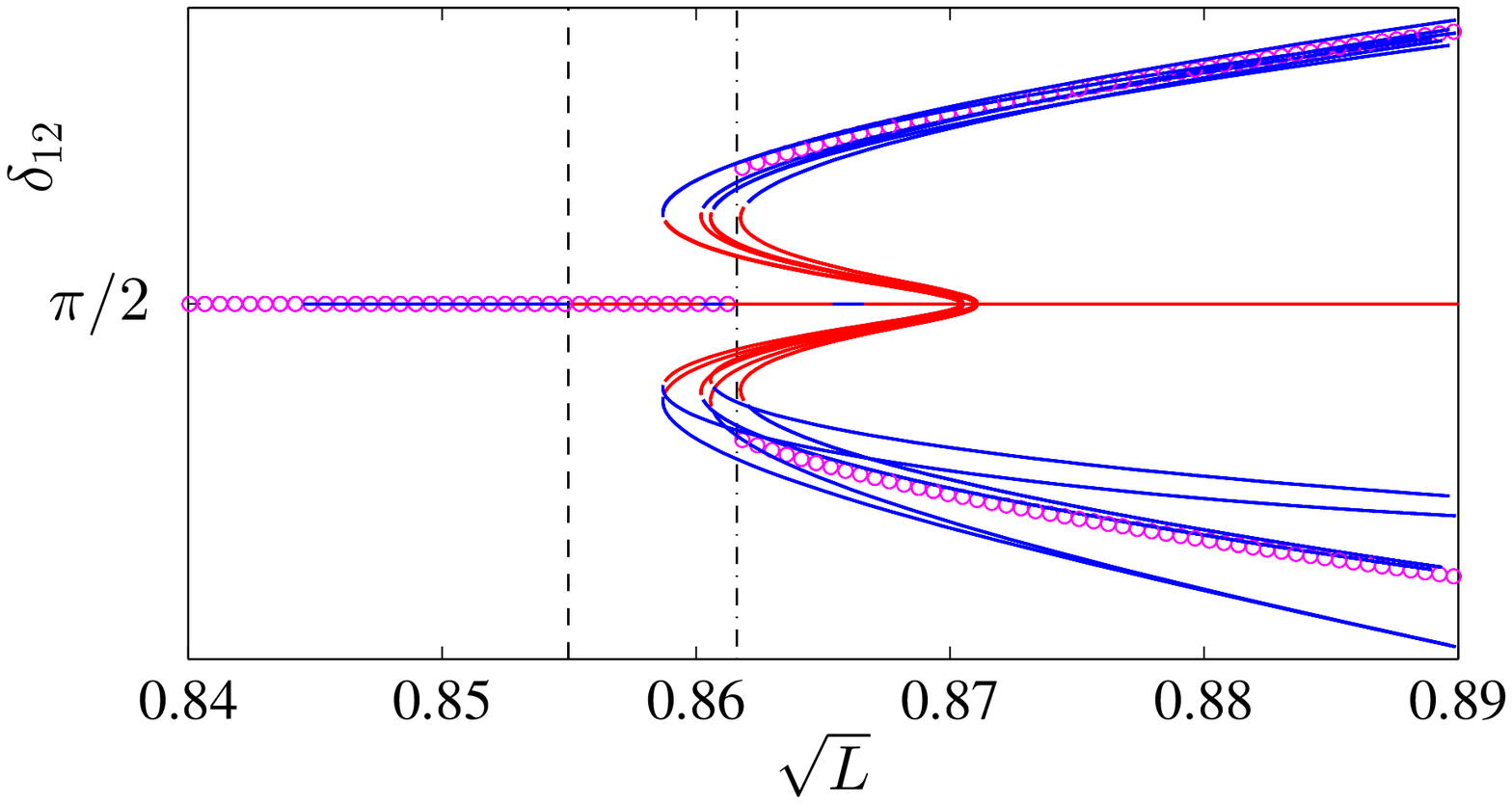}
\end{center}
\caption{(Color online).
Bifurcation scenario for the relative angles between vortices $\delta_{12}$
as a function of the (square root) of the angular momentum
for $N=4$ vortices.
The data is the same as Fig.~\ref{N4} and the notation is also the same.
The top panel is the full view while the bottom panel depicts a zoomed in
version around the bifurcation region. Notice that between the transition
from the square to the rhombus (dashed vertical line) and that from
the rhombus to the general quadrilateral (dash-dotted vertical
line), no modification is
noticed by looking solely at the angles, as the rhombic configuration
maintains $\delta_{12}= \pi/2$.}
\label{N4pi}
\end{figure}

\begin{figure*}[htbp]
\begin{center}
\includegraphics[width=8cm]{\rootfig 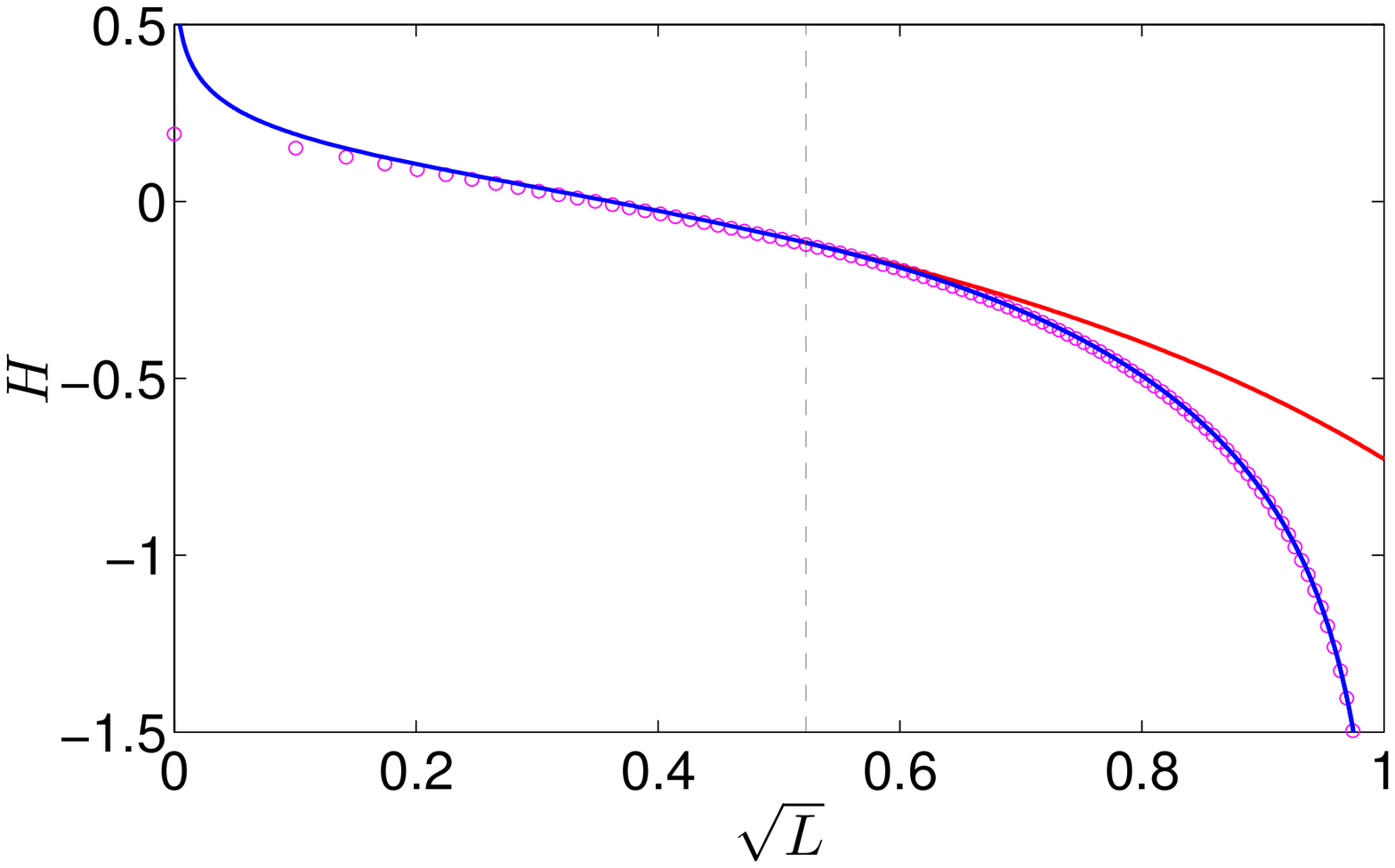}
\includegraphics[width=8cm]{\rootfig 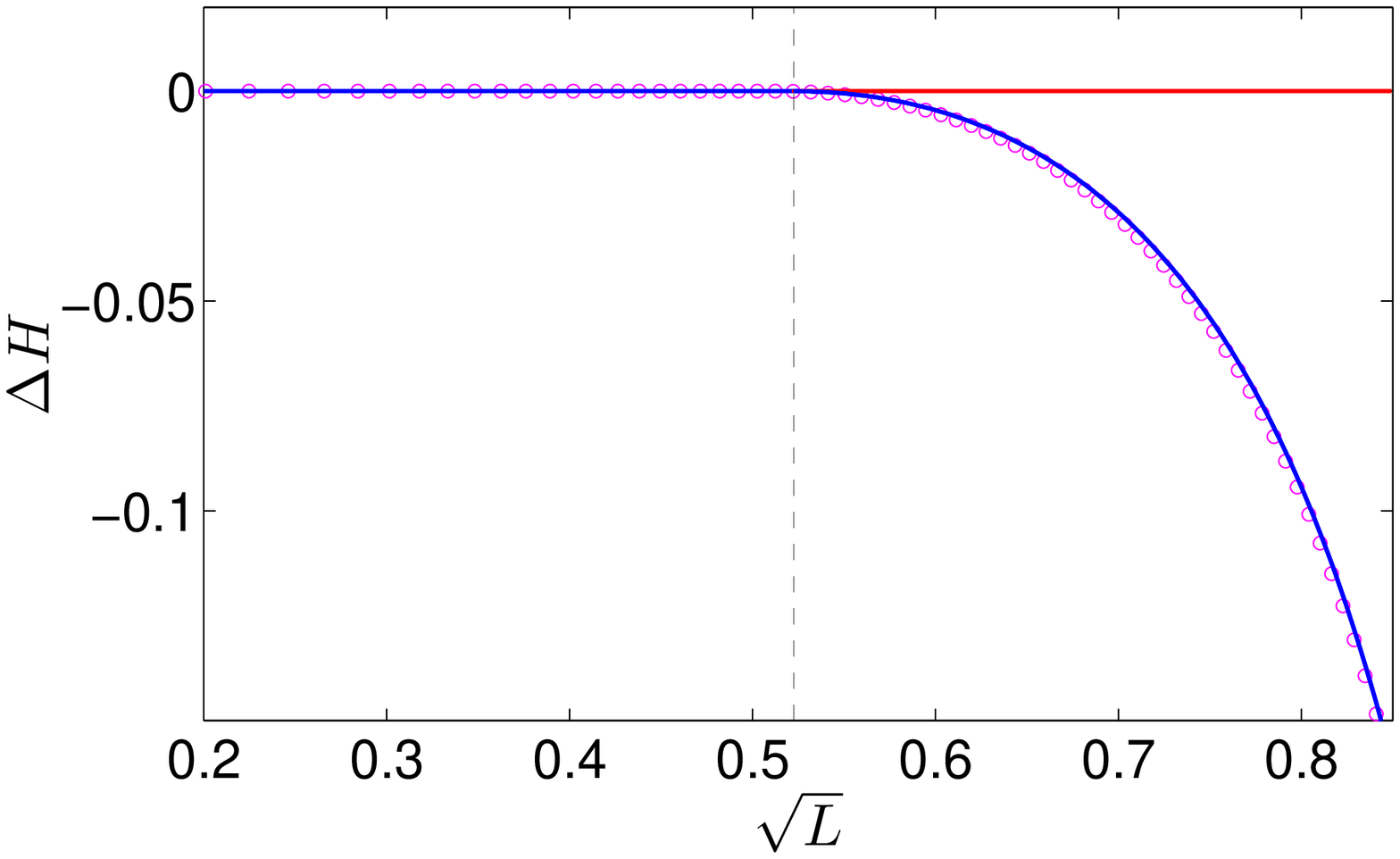}\\[1.0ex]
\includegraphics[width=8cm]{\rootfig 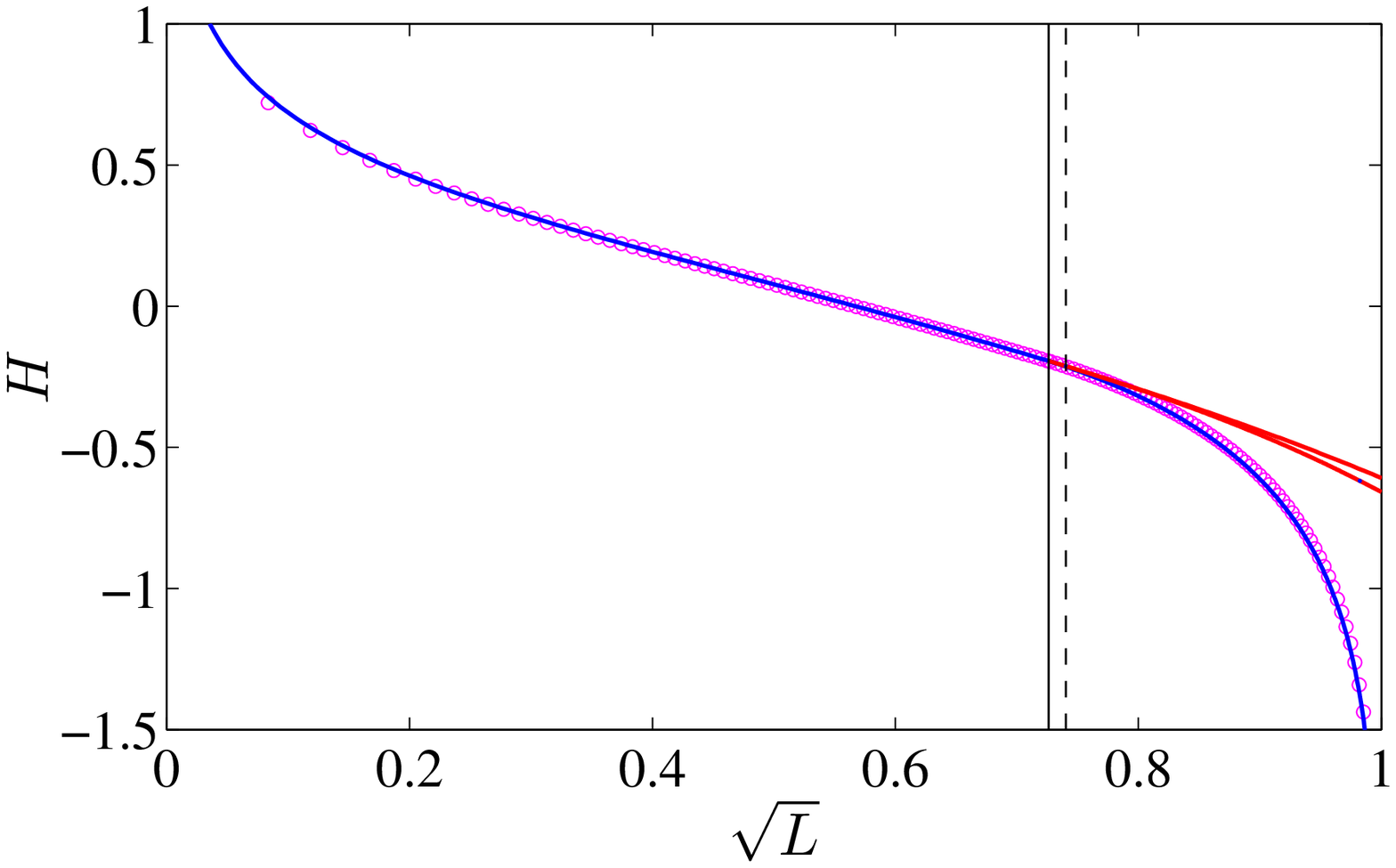}
\includegraphics[width=8cm]{\rootfig 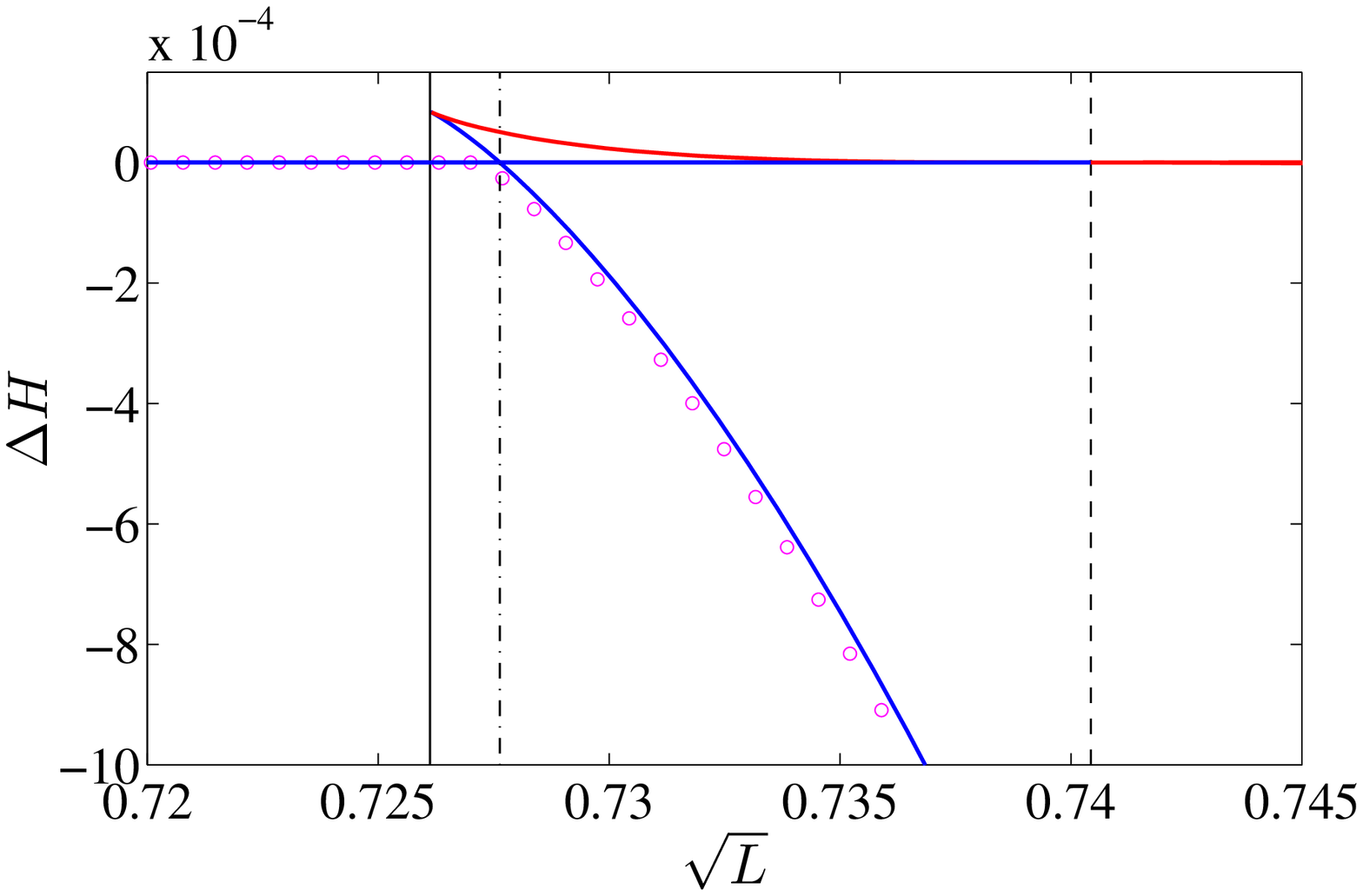}\\[1.0ex]
\includegraphics[width=8cm]{\rootfig 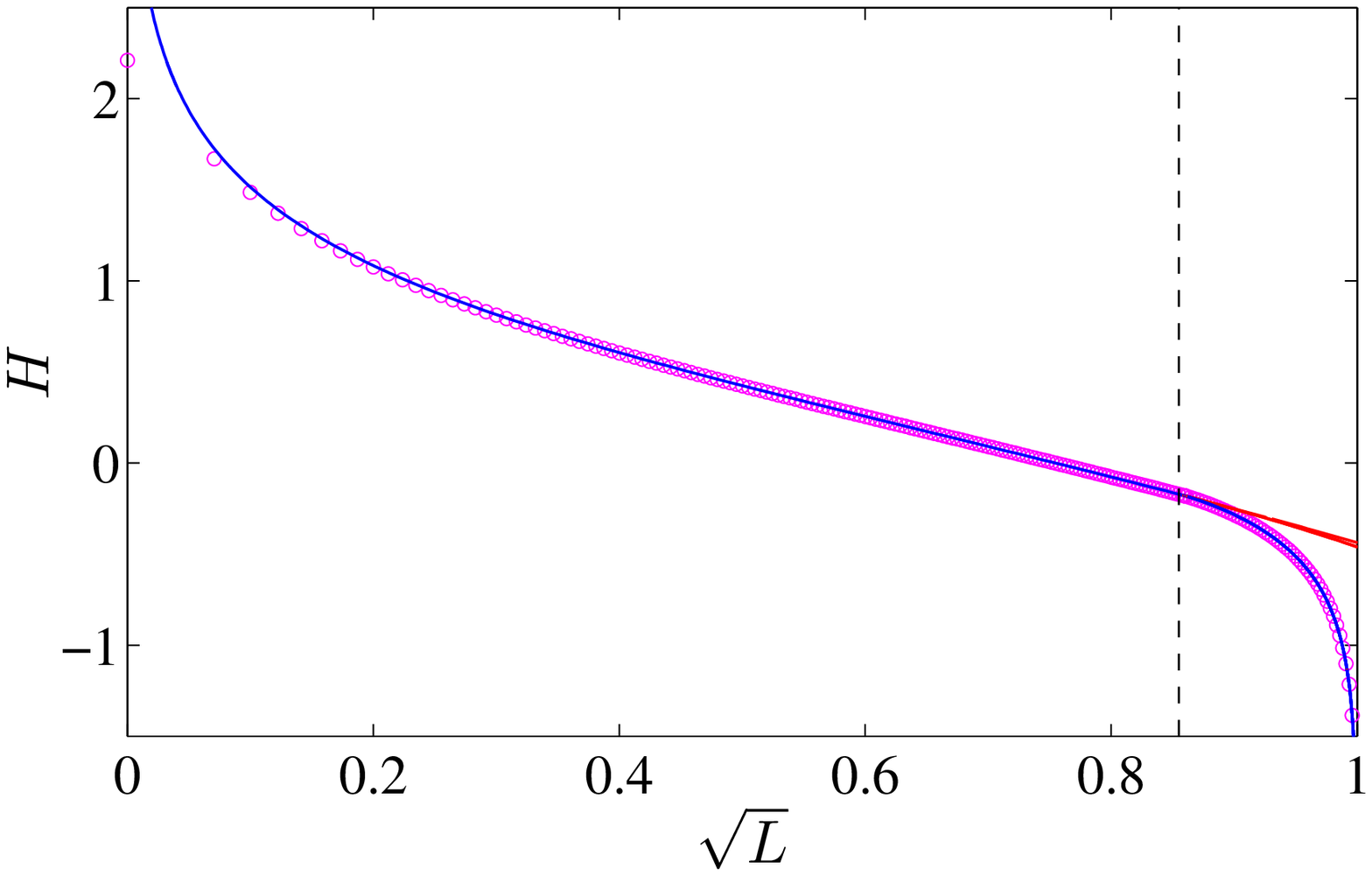}
\includegraphics[width=8cm]{\rootfig 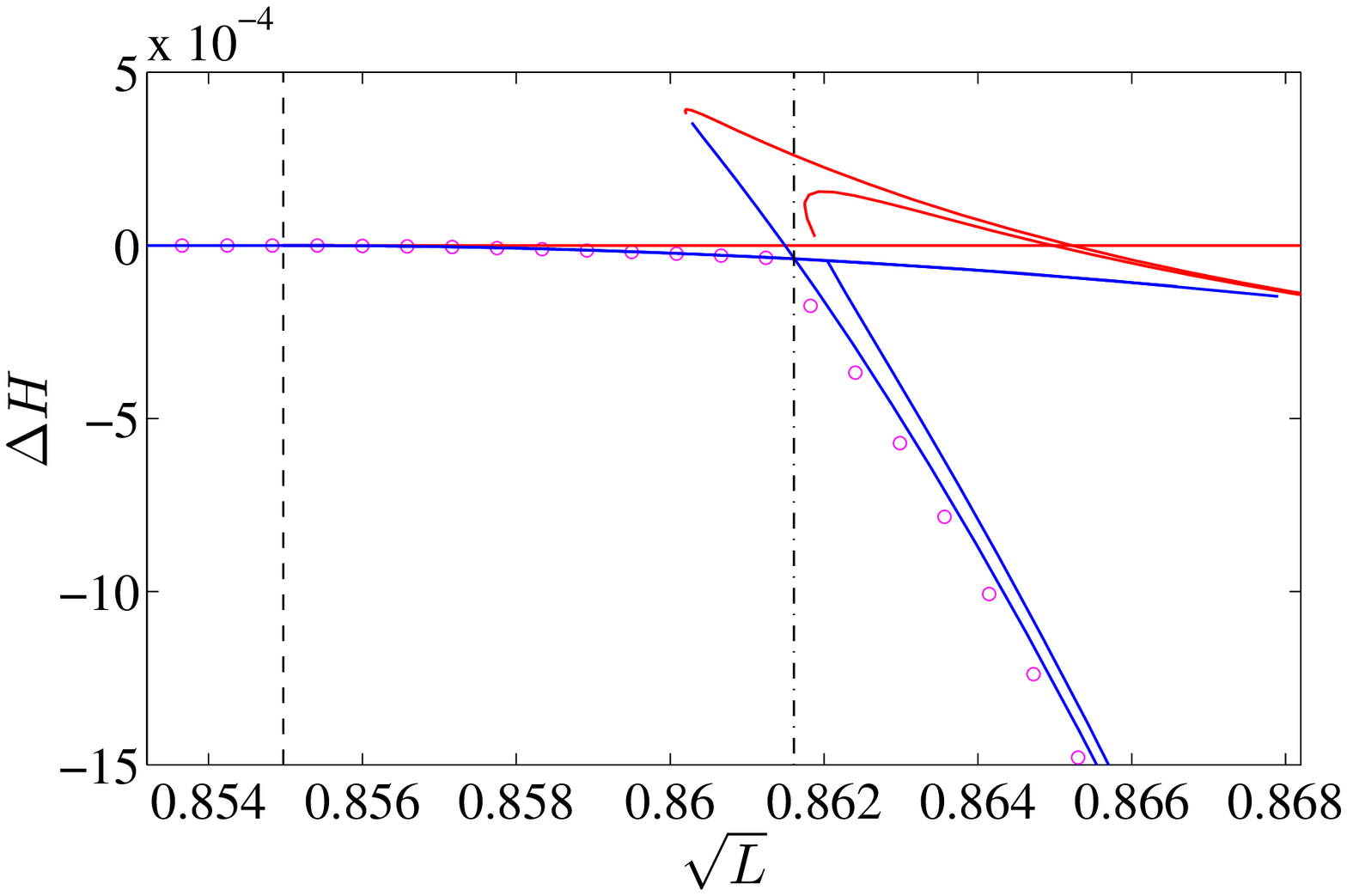}\\[1.0ex]
\end{center}
\caption{(Color online).
Energy corresponding to all the bifurcation branches
presented in the previous plots together with the
energy computed from the MC simulations. As before,
the bifurcation branches are depicted by a blue (red)
line for stable (unstable) branches and the MC simulations
are depicted with the magenta circles.
The first, second and third rows of panels correspond,
respectively, to the $N=2$, $3$ and $4$ cases.
The left panels present the energy as computed from
Eq.~(\ref{H}) over the entire range $0\leq L \leq 1$,
while the right panels depict the corresponding energy
difference $\Delta H$ between each configuration and
the corresponding symmetric configuration for each
value of $L$ (see text for more details).
}
\label{fig:H}
\end{figure*}

In addition to the square configuration, we have again sought
the possibility of unveiling analytically reduced symmetry solutions.
An example that we have been able to identify in this case is a
{\it rhombic} configuration with $r_1=r_3$ and $r_2=r_4$ in which case
still all the $\delta_{ij}=\pi/2$. For this configuration we have
been able to find that it consists of two longer and two shorter
segments $r_1$ and $r_2$ such that $L=2 (r_1^2 + r_2^2)$ and
$r_2= \sqrt{c (1 - r_1^2)/ (2 r_1^2 + c (r_1^2 -1))}$. It is then
straightforward to observe that this configuration ``collides''
with the square branch (i.e., $r_2=r_1$) exactly at
$r= r_{cr,1}^{(4)}$ which is precisely where the square configuration
loses its stability through the zero-crossing of the
frequency $\omega_1$.
From the above and since this solution exists only
above this critical point, it can be inferred that the primary
instability of the square configuration leads to a supercritical
pitchfork bifurcation that, in turn, results in the emergence of
the rhombic state.
This is confirmed in Fig.~\ref{N4} where the location of
this primary bifurcation indeed occurs at $r_{cr,1}^{(4)}$,
see dashed vertical line.

Numerically, we indeed observe this destabilization and the
corresponding symmetry breaking bifurcation which is
depicted in Fig.~\ref{N4}.
In particular,
a destabilization event
for $c=0.1$ clearly arises at $\sqrt{L}=0.862$
in the Monte-Carlo (see vertical dash-dotted line), while the
corresponding analytical prediction is at $(L_{cr,1}^{(4)})^{1/2}=0.855$
(see vertical dashed line). It is particularly
interesting that close inspection of Fig.~\ref{N4} reveals for
a few points between $0.855$ and $0.862$ the transition from the
square to the rhombi (although the growth rate of the associated
instability in this interval is apparently so weak that the MC
may still converge to the squares for some values of $\sqrt{L}$
within this interval).
On the other hand, we also depict the relevant bifurcation for the
relative angles $\delta_{ij}$ in Fig.~\ref{N4pi}.
Remarkably but also naturally, between $0.855$ and $0.862$,
and while the radii reveal (at least partially) the transition
from the squares to the rhombi, $\delta_{12}$ remains invariant
at $\pi/2$, as it is shared by both configurations. Hence, it is
clear that one cannot use solely the radii or solely
the relative angles, but a careful inspection of both unveils
the full picture of configurational transitions.
%
%
For slightly higher values of the angular momentum,
i.e., for $\sqrt{L} > 0.8626$, the Monte-Carlo jumps to another
configuration which in this case does not appear to have any
definite symmetry. While the
relative angles $\delta_{ij}$, as discussed above,
were unable to ``discern'' the first transition
(the supercritical pitchfork from the square to the rhombus),
nevertheless,
they  clearly distinguish the second transition, whereafter
none of the angles is equal to $\pi/2$.

It should be clear at this point that, as in the $N=3$ case, in the
$N=4$ examples as well, the dynamical picture offers a particularly
useful complementing view which corroborates in an insightful manner
the results of the Monte-Carlo approach. In particular, we clearly observe
the transition from the square to the rhombus.
The latter state, however,
is apparently the ground state of the $N=4$ system only
for a small interval of angular momenta. This
is because already for $\sqrt{L}=0.859$ a pair of asymmetric
(so-called irregular) quadrilaterals of vortices arise with
unequal sides, yet rigidly rotating around the center of the
trap. Remarkably, one of the highly asymmetric configurations that
arise in this saddle-center bifurcation is dynamically stable
and it is that one that becomes the global energy minimum beyond
the second critical point, namely $\sqrt{L}=0.8626$. For these
quadrilaterals it can be seen that approximately
$r_1=r_3$, $r_2$ is close to $r_1,r_3$ but clearly not equal
and $r_4$ much larger (rotated versions of such quadrilaterals
also obviously exist from symmetry). Interestingly, the
dynamical picture reveals one more feature, namely that such
quadrilaterals collide via a sub-critical pitchfork
with the rhombic configuration for  $L^{1/2}=0.87$.
I.e., the full dynamics and stability picture is far more
complicated, involving a series of bifurcations, a super-critical
and a sub-critical pitchfork, as well as a saddle-center bifurcation, yet
again the combination of the Monte-Carlo method and the
bifurcation analysis yields a complete understanding of the
system's ground state features.

Finally, in order to more precisely illustrate the
feature that the MC simulation is indeed
converging to the stable state with minimum energy,
we have followed the Hamiltonian (\ref{H}) as the angular
momentum is varied. The results for $N=2$, $3$ and $4$ are
depicted in Fig.~\ref{fig:H}. The left column on the panels
corresponds to the total energy as given by Eq.~(\ref{H}),
while the right panels depict zoomed in versions for
the energy difference $\Delta H$ between the configuration at hand
and the symmetric state. Namely, we define
$\Delta H = H - H_0$, where $H$ is computed using Eq.~(\ref{H})
for each configuration and $H_0=H(r_i=r^*,\delta_{i,i \pm 1}=\delta^*)$
where $(r^*,\delta^*)$ correspond to the radius and relative angle
for a {\em symmetric} polygonal configuration. Namely, for $N$
vortices these correspond to $r^* = \sqrt{L/N}$
and $\delta^*=2\pi/N$.
In the case of $N=2$, the picture is very clear: as soon as
the asymmetric anti-diametric configuration emerges, the MC converges to it.
For $N=3$, the emergence of the asymmetric isosceles states occurs
well before the destabilization of the equilateral, co-rotating
triangle branch. Nevertheless, very shortly after
the saddle-center bifurcation of this new branch (cf. the
vertical solid with the vertical dash-dotted line in the
middle right panel Fig.~\ref{fig:H}), the MC jumps to the newly
emergent, asymmetric branch {\it as soon as} the latter
becomes energetically favorable, yet well before
the instability of the equilateral branch (occurring at the location
of the vertical dashed line).
The case of $N=4$ is more complex.
Here we can see that once the square configuration destabilizes
towards the rhombic one (vertical dashed line),
the MC follows the rhombi until very shortly after the
emergence of the irregular quadrilateral branch; the latter, is generated 
by the saddle-center bifurcation, and 
acquires lower energy than the rhombi (vertical dash-dotted line).
Immediately thereafter, the MC approach traces this and jumps to it.
It is clear from the results presented in Fig.~\ref{fig:H} that the MC
simulations indeed converge, for a given $L$ to the stable state which has
the lowest energy among the different vortex configurations.

\begin{figure}[htbp]
\begin{center}
\includegraphics[width=8cm]{\rootfig 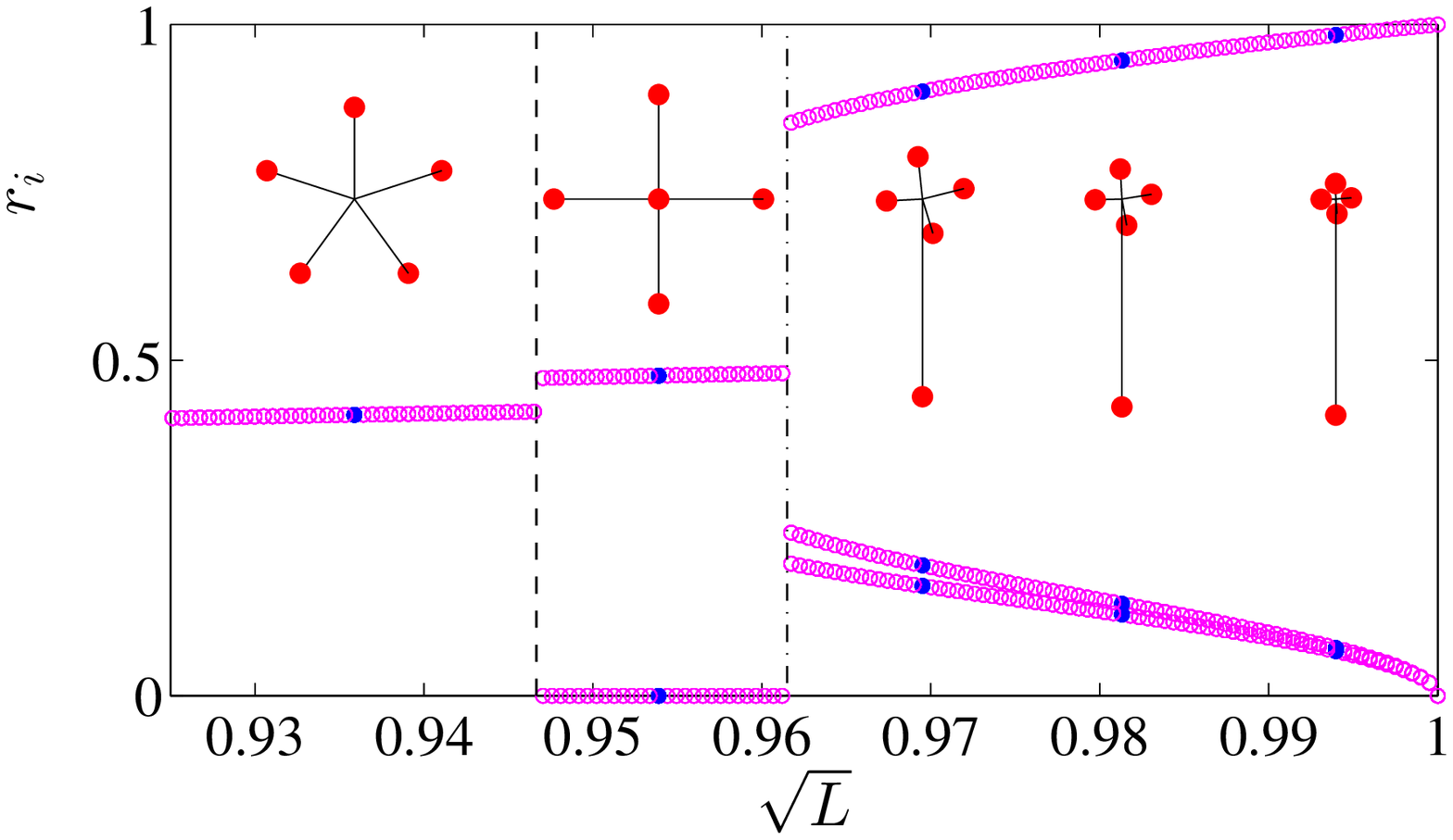}
\includegraphics[width=8cm]{\rootfig 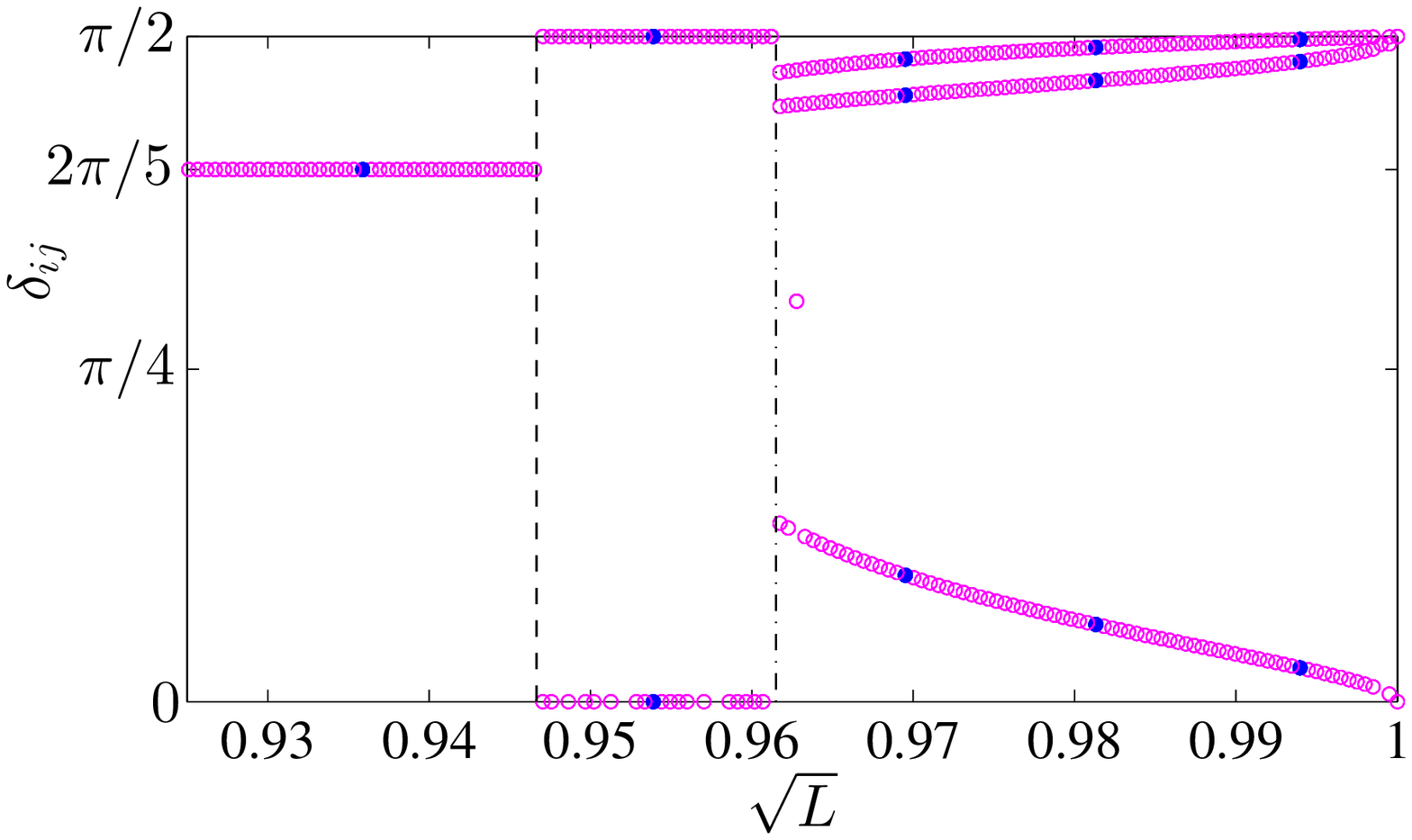}
\end{center}
\caption{(Color online)
Monte-Carlo results for $N=5$ vortices. The top and
bottom panel depict, respectively, the MC results for the radii ($r_i$)
and the relative angles ($\delta_{ij}$).
Two transitions are observed.
The first transition at $L^{1/2}=0.9467$ (location depicted by the
vertical dashed line) indicates where the configuration
with a single vortex at the center turns into the ground state of
the system.
The second transition observed at $L^{1/2}=0.9615$ (location
depicted by the vertical dash-dotted line)
indicates where the asymmetric configuration
(a tight cluster of 4 vortices plus a single
vortex further away from the center)
turns into the ground state of the system.
The filled (blue) dots along the different branches
indicate the locations of the displayed vortex configurations
in the top panel.
}
\label{N5r}
\end{figure}

\subsection{The N=5 Vortex Case}

For the case of $N=5$, the relevant calculations both analytical
and numerical become, arguably, very complex. Nevertheless, we have
still been able to analyze the stability of the pentagon configuration
with $r=r_i$, $L=5 r^2$ and $\delta_{ij}=2 \pi/5$. Such configurations will
be stable, remarkably, until the same principal critical point as
were $N=3$ and $N=4$ polygons, namely
$r^2= (r_{cr,1}^{(5)})^2 \equiv {\sqrt{c}}/({\sqrt{c} + \sqrt{2}})$,
although, of course this corresponds to a higher angular momentum
for this case, namely $L_{cr,1}^{(5)}=5(r_{cr,1}^{(5)})^2$.
On the other hand, we have also been
able to identify a second critical point which arises at
$r^2= (r_{cr,2}^{(5)})^2 \equiv {\sqrt{c}}/({\sqrt{c} + 1})$,
i.e., at a higher radius.
This first critical point occurs analytically at $\sqrt{L}=0.956$,
while the Monte-Carlo numerically appears to deviate from the
pentagon configuration for the earlier value of $\sqrt{L}=0.9467$.
However, as is clear from the Monte-Carlo results of Fig.~\ref{N5r},
the bifurcation occurring at this point cannot be a supercritical
pitchfork one, given the sizeable ``jump'' of the values of
the $r_i$'s occurring at this point.
Under close inspection, this first transition
captured by the MC simulations corresponds to
a value for the angular momentum where another, independent
configuration branch becomes the ground state of the system.
In fact, apparently, for values of the angular momentum in
$0.9467<L^{1/2}<0.9615$, the configuration bearing a single
vortex at the center surrounded by a square of vortices has
less energy than the pentagon.
For larger values of the angular momentum,
$L^{1/2}>0.9615$, the asymmetric configuration bearing
a tight cluster of four vortices near the origin and a
single vortex further away from the center, corresponds
to the lowest energy configuration of the system.
Identifying the conditions for the existence of windows
where the configuration with a single vortex at the center
with a polygon of $N-1$ vortices around it is more energetically
favorable than a polygon of $N$ vortices remains an
interesting open problem for future work.

\section{Conclusions and Future Work}

In the present work, we have used a combination of analytical
and numerical techniques to shed light on the (already fairly
complex for small number of vortices) possible solutions
and associated bifurcations of co-rotating vortices in
atomic Bose-Einstein condensates. Building on the earlier
establishment of a relevant model through comparisons
with experimental results, e.g., in Refs.~\cite{middel_pra11,corot_prl},
we developed both a Monte-Carlo approach targeting the lowest
energy states and an AUTO-based dynamical systems approach
attempting to infer the relevant solutions and their pitchfork
and saddle-center bifurcations into existence/termination.
By corroborating the two techniques and using the angular
momentum as a parameter, and the energy as well as the
vortex positions as diagnostics, we
were able to provide a full picture of how two rigidly
rotating vortices remain anti-diametric but become
asymmetric, three rigidly rotating vortices prefer to
be in an isosceles rather than equilateral triangle, while
four turn from squares to rhombi and from there to irregular
quadrilaterals. All these transitions have been quantified
as a function of increasing angular momenta and ultimately
result from the competition of the two energetic
contributions in Eq.~(\ref{H}), namely the precession of each
vortex due to the
trap and the pairwise interaction between the vortices.  Whenever possible
the numerical observations have been complemented by analytical
solutions (e.g. identifying the destabilization points of
symmetric configurations, or analytically characterizing the
bifurcating solutions such as isosceles triangles and rhombi).

Nevertheless, naturally many open questions still remain
and the system clearly merits further investigation.
As an appetizer towards that direction, we presented the
calculation of $N=5$, indicating a clearly subcritical
event that must be leading to the destabilization of the
pentagons. Our computational approaches have natural
limitations that arise both for the dynamical systems
AUTO-based analysis and for the Monte-Carlo efficient
ground state tracking method. We now briefly discuss these
limitations and present a view towards overcoming them
in the future which would indeed enable a systematic
categorization of larger vortex particle clouds.

On the one hand, the AUTO calculation is extremely useful
in identifying the relevant bifurcations, but given that it
tracks the different branches of solutions, it provides a progressively
more complex and difficult to parse picture as $N$ is increased. Hence, it is
necessary to use multiple and different suitably chosen diagnostics
in order to be able to systematically scale up the picture to
cases of larger $N$. It would be particularly meaningful of a task
to try to develop such diagnostics and it is part of our currently ongoing
effort. On the other hand, the Monte-Carlo approach suffers from
a different limitation most notably the divergence of vortex
precessional frequencies (and logarithmic associated single vortex
energy contributions) when $r_i \rightarrow 1$. It is precisely
for that reason that we have confined our consideration on the
MC side to $L < 1$. It naturally turns out that when $L \geq 1$,
the energy minimization can be trivially (but meaninglessly,
as far as the physical problem is concerned) be realized
by means of one (or more) of the $r_i \rightarrow 1$ and
hence $H \rightarrow -\infty$. Hence, it is of paramount
importance in that regard to amend the ``pathological''
precessional frequency expression with one that more
accurately predicts the $r \rightarrow 1$ regime in
comparison to the partial differential equation (PDE) (see also
the relevant partial disparity in Fig. 1a of Ref.~\cite{corot_prl}, for
vortices at distances very proximal to the TF radius).
A combination of variants of the above tools devoid of these
technical limitations (for small and intermediate $N$)
and possible intriguing tools from PDE theory about vortex
``densities'' (for large $N$) in the spirit e.g. of the
recent work of Ref.~\cite{theo} can provide valuable insights
for future studies of vortices, but also of other
types of solitonic populations, such as dark solitons
in 1D or vortex rings in 3D BECs~\cite{ourexpo}. Such
studies are currently in progress and will be reported in
future publications.

Finally, it is important to note that the results we
present in this manuscript are based on the assumption of
an axially symmetric trapping potential. If one relaxes
this symmetry and considers different trapping strengths
along the longitudinal directions, the vortex precession rate
has to be adjusted and depends on the angular position
of the vortex with respect to the trapping axes
\cite{Fetter_Svidzinsky_2000}; see also Ref.~\cite{middel_anis}
for multi-vortex settings. The dynamics for
asymmetric trapping is much richer than the one presented here
and will be showcased in a future publication.

\acknowledgments R.C.G. and P.G.K. gratefully acknowledge
support from the National Science Foundation under grant
DMS-0806762. P.G.K. also acknowledges support from
CMMI-1000337, from the Alexander von Humboldt Foundation,
from the Binational Science Foundation under grant 2010239
and from the US AFOSR under grant FA9550-12-1-0332.
The work of F.K.D. and
D.J.F. was partially supported by the Special Account for Research Grants
of the University of Athens.

\end{document}